%% file: turtledraft.tex
\documentclass[11pt,a4paper]{article}
\pdfoutput=1
\usepackage{jheppub}
\usepackage[utf8]{inputenc}

\usepackage{color}
\usepackage{amsmath}
\usepackage{pifont}
\usepackage{bbold}

\usepackage{bbm}
\usepackage{verbatim}   
\usepackage{subfigure}
\usepackage{acronym}

\usepackage{ams fonts}
\usepackage{amssymb}
\usepackage{mathrsfs}
\usepackage{slashed}
\usepackage{xfrac}

\usepackage{soul}

\usepackage{caption}
\usepackage{tikz}
\usetikzlibrary{decorations.pathreplacing}

\usepackage{scalerel}

\usepackage{subfiles}
\usepackage{cancel}
\usepackage{units}

\newcommand{\fig}[1]{Fig.~\ref{fig:#1}}
\newcommand{\tab}[1]{Table~\ref{tab:#1}}
\newcommand{\tmnt}{\scaleto{\rm TMNT}{4pt}}
\newcommand{\twin}{\scaleto{\rm TH}{4pt}}
\newcommand{\sm}{\scaleto{\rm SM}{4pt}}
\newcommand{\soft}{\rm soft}
\newcommand{\hard}{\rm hard}

\title{Twin Turtles}
\author[a,d]{Pouya Asadi}
\emailAdd{asadi@physics.rutgers.edu}
\author[b,d]{Nathaniel Craig,}
\emailAdd{ncraig@physics.ucsb.edu}
\author[c,d]{and Ying-Ying Li}
\emailAdd{ylict@connect.ust.hk}

\affiliation[a]{NHETC, Department of Physics and Astronomy, Rutgers University, Piscataway, NJ 08854, USA}
\affiliation[b]{Department of Physics, University of California, Santa Barbara, CA 93106, USA}
\affiliation[c]{Department of Physics, The Hong Kong University of Science and Technology, Clear Water Bay, Kowloon, Hong Kong S.A.R., P.R.C}
\affiliation[d]{Kavli Institute for Theoretical Physics, Santa Barbara, CA 93106, USA}

\abstract{
We present an ultraviolet extension of the Twin Higgs in which the radial mode of twin symmetry breaking is itself a pseudo-goldstone boson. This ``turtle'' structure raises the scale of new colored particles in exchange for additional states in the Higgs sector, making multiple Higgs-like scalars the definitive signature of naturalness in this context. We explore the parametrics and phenomenology of a concrete Twin Turtle model and demonstrate its robustness in two different supersymmetric completions. Along the way, we also introduce a new mechanism for inducing hard twin symmetry-breaking quartics via soft supersymmetry breaking. 
}

\begin{document} 
\maketitle
\flushbottom

\section{Introduction}
\label{sec:intro}

\subfile{introduction}

\section{Tunings in Twins and Turtles}
\label{sec:tuning}

\subfile{tuning}

\section{A Bottom-Up Twin Turtle}
\label{sec:bottomup}

\subfile{bottomup}

\section{A Top-Down Twin Turtle}
\label{sec:topdown}
Thus far, we have treated the Twin Turtle purely as an effective theory up to cutoffs $\Lambda_t$ and $\Lambda_\rho$ in the top sector and Higgs sector. We now turn to the construction of supersymmetric UV completions of this framework, which demonstrate the robustness of the mechanism and validate several parametric choices made in our exploration of the bottom-up theory. 

The leading quartic couplings in the TMNT model may be generated via either $F$-term or $D$-term quartics in the supersymmetric UV completion \cite{Chang:2006ra, Falkowski:2006qq, Craig:2013fga,Batra:2003nj,Badziak:2017syq}, and we explore both options. Along the way, we present a mechanism for generating the hard-breaking quartics purely from soft breakings of $\mathbb{Z}_2$ symmetries. This has the advantage of reducing UV sensitivity and restoring all of the relevant $\mathbb{Z}_2$ symmetries as exact symmetries of the dimensionless couplings in the theory. While we make use of the mechanism for the TMNT, it is generally applicable to supersymmetric UV completions of Twin Higgs models employing hard-breaking quartics.

\subsection{An $F$-term Model}
\label{subsec:fterm}
\subfile{topdownF}

\subsection{A $D$-term Model}
\label{subsec:dterm}

\subfile{topdownD}

\section{Conclusions}
\label{sec:conclusion}

\subfile{conclusion}

\acknowledgments

We thank David Curtin, Simon Knapen, Seth Koren, Lingfeng Li, Tao Liu, Diego Redigolo, Scott Thomas, Lian-Tao Wang, and Yue Zhao for helpful conversations. NC also thanks Nima Arkani-Hamed for reminding him of \cite{Batra:2004ah} in the first place, and Tim Cohen for enlightening conversations about TMNTs in general. NC is supported in part by the US Department of Energy under the Early Career Award DE-SC0014129 and the Cottrell Scholar Program through the Research Corporation for Science Advancement. PA is supported by DOE grant DE-SC0010008. YYL is supported by the Hong Kong PhD Fellowship Scheme (HKPFS). PA and YYL thank the Kavli Institute for Theoretical Physics for the award of a graduate visiting fellowship, provided through Simons Foundation Grant No.~216179 and Gordon and Betty Moore Foundation Grant No.~4310. We thank the Kavli Institute of Theoretical Physics for hospitality during the inception of this work, and corresponding support from the National Science Foundation under Grant No. NSF PHY-1748958. 

\bibliography{biblio}

\end{document}

%% file: introduction.tex
The electroweak hierarchy problem is one of the primary motivators of physics beyond the Standard Model, with most proposed solutions predicting an abundance of new states near the weak scale. But the discovery of an apparently-elementary Higgs at the LHC, coupled with the non-discovery of any additional particles, raises a pressing question: if the hierarchy problem is solved by new physics, {\it where is everyone?} Attempts to answer this question have given rise to a variety of new approaches to the hierarchy problem, including selection of the weak scale by cosmological dynamics \cite{Graham:2015cka} and stabilization of the weak scale by discrete symmetries, rather than continuous ones \cite{Chacko:2005pe, Burdman:2006tz, Craig:2014aea}. Typically these new approaches are at most successful in stabilizing the Higgs mass up to a cutoff intermediate between the weak scale and the Planck scale, and still require one of the two conventional solutions (supersymmetry or a low cutoff, possibly from compositeness \cite{Kaplan:1983fs,Kaplan:1983sm}) to cover the remaining ground between the intermediate cutoff and the Planck scale.

The problem is particularly acute in models of `neutral naturalness' that exploit discrete symmetries. While the lowest-lying states stabilizing the weak scale in these models are neutral under the Standard Model, they only succeed in raising the cutoff (at fixed fine-tuning) by a modest amount of order $g_* / g_{SM}$ \cite{Contino:2017moj}, where $g_*$ is the coupling strength of new dynamics associated with the discrete symmetry and $g_{SM}$ is a relevant Standard Model coupling (typically the top yukawa or Higgs quartic, depending on the model). Thus supersymmetry or compositeness must still enter at a scale only modestly higher than before, albeit high enough to put typical SM-charged partner particles (just) outside the reach of the LHC \cite{Falkowski:2006qq, Chang:2006ra, Craig:2013fga, Geller:2014kta, Barbieri:2015lqa, Low:2015nqa, Katz:2016wtw, Badziak:2017syq}.

The success of global symmetries in raising the cutoff of a natural Higgs sector has long hinted at a more ambitious possibility: that the ultimate cutoff could be raised further by successive patterns of global symmetry breaking, in which the scalar spontaneously breaking a global symmetry is itself a pseudo Nambu-Goldstone boson (pNGB). Theories in which the scale associated with the breaking of a global symmetry is stabilized by the breaking of another global symmetry are referred to as `turtles' \cite{Kaplan:2004cr}, borrowed from Hindu mythology by way of Hawking \cite{hawkingbrief}. Turtles involving two successive Little Higgs models were constructed in \cite{Kaplan:2004cr,Batra:2004ah}, pushing the ultimate cutoff of the perturbative, nonsupersymmetric construction up to $\sim 100$ TeV. Further raising the cutoff by additional symmetry-breaking structures was ultimately limited by the geometric increase in the number of fields at each level of the turtle, which reduce the corresponding expansion parameter. The possibility of constructing novel multi-level turtles with perturbative 't Hooft coupling was sketched in \cite{Batra:2004ah} but has yet to be exploited further.

In this paper we revive the turtle mechanism in the context of the simplest model of neutral naturalness, the Twin Higgs \cite{Chacko:2005pe}. There are some natural advantages to this approach, as the use of discrete symmetries (rather than continuous ones) reduces the multiplicity of fields transforming under a given gauge group at each level of the turtle. But the introduction of twin turtles is also strongly motivated by LHC null results, in that it promises to further raise the scale at which new particles carrying Standard Model QCD quantum numbers must appear. Whereas the pre-LHC goal of the original turtle models was to push off the scale of supersymmetry or compositeness as far as possible by introducing a host of Standard Model-charged states, our objective here is more modest: to raise the scale associated with new charged states as far as possible, whether or not they are associated with supersymmetry or compositeness. The irreducible scale at which new colored particles appear in a natural completion of the weak scale is of particular relevance to the physics case for future colliders (see, e.g.~\cite{Curtin:2015bka}).

However, there is an immediate obstruction to naively concatenating patterns of global symmetry breaking in simple models of neutral naturalness such as the Twin Higgs: raising the cutoff with turtles is only manifestly sensible in Little Higgs-like theories where there is no additional tuning penalty associated with vacuum alignment. In conventional composite Higgs models or the Twin Higgs, the separation between two scales of spontaneous symmetry breaking $f_2 \gg f_1$ is typically associated with a tuning of order $f_2^2 / f_1^2$. In this case the concatenation of $n$ successive symmetry breaking scales would accumulate a tuning of order $(f_n^2/ f_{n-1}^2) \times \dots \times (f_2^2/f_1^2) = f_n^2/f_1^2$, no less tuned (and significantly more baroque) than a two-scale theory with large separation of scales.

Progress is possible if the tuning associated with vacuum alignment can be reduced. For Twin Higgs models, there are three possibilities: tadpoles \cite{Harnik:2016koz}; collective symmetry breaking {\it \`{a} la} Little Higgs \cite{Chacko:2005vw}; or hard breaking of the discrete symmetry \cite{Chang:2006ra}. Here we will focus on the latter option, which is relatively straightforward to implement and entails fairly minimal field content.\footnote{By using hard breaking to eliminate vacuum alignment tuning, we depart from the original spirit of \cite{Kaplan:2004cr}, which aimed to reduce all cutoff dependence to the greatest possible extent; here our focus is primarily on raising the scale associated with the top sector.} Indeed, hard symmetry breaking is particularly well-suited to turtle extensions of the Twin Higgs. In simple Twin Higgs models, the size of the hard-breaking quartic -- which in turn dictates the reduction in vacuum alignment tuning -- is bounded by the known small size of the Standard Model-like Higgs quartic. But in turtle extensions of the Twin Higgs, hard breaking can be used to separate intermediate symmetry breaking scales, for which the associated quartics are unfixed. 

This motivates us to study Twin Turtle models in which the vacuum alignment separating the weak scale from the first scale of global symmetry breaking is achieved by the soft breaking of a discrete symmetry, while the separation of this first scale from higher scales of global symmetry breaking is achieved by the hard breaking of discrete symmetries. In what follows, we will refer to this particular construction as the Twin Maximally Natural Turtle (TMNT). Ultimately, we will find that this particular structure is capable of parametrically increasing the scale associated with the appearance of new colored particles relative to the Twin Higgs, albeit at the cost of lowering the scale associated with the appearance of new particles in the Higgs sector. In this respect, the TMNT represents a class of models in which additional Higgs-like scalars become the definitive signature of naturalness.\footnote{For other Twin Higgs-inspired models in which additional Higgs bosons improve electroweak naturalness, albeit in different ways, see \cite{Craig:2014roa, Yu:2016swa}.}

Thanks to the particular choice of symmetry breaking patterns, the Twin Turtle models explored here exhibit improved tuning by typical metrics such as the Barbieri-Giudice fine-tuning measure \cite{Barbieri:1987fn}, but are perhaps unlikely to satisfy the Potter Stewart measure \cite{Stewart:1964}. It is not our goal to satisfy every possible measure of tuning, but rather to understand the extent to which the scale of new colored particles may be delayed in natural completions of the weak scale, and to additionally lay the groundwork for renewed exploration of turtle models motivated by LHC null results. While these constructions demonstrate an improvement in tuning at the inarguable cost of complexity, they illustrate new parametric possibilities that nature may find a more elegant means to saturate.

The paper is organized as follows: In Section \ref{sec:tuning} we discuss the parametrics of fine tuning in various incarnations of the Twin Higgs and its turtled extensions, motivating the particular combination of hard- and soft-breaking terms that characterize the Twin Maximally Natural Turtle. We then explore the parametrics and phenomenology of the TMNT in Section \ref{sec:bottomup} from the perspective of an effective theory with unspecified physics at the cutoff. In \ref{sec:topdown} we construct two supersymmetric UV completions of the TMNT, which serve to both justify the symmetry-breaking pattern 
and validate the tuning expectations of the effective theory. We explore more speculative directions for twin turtle model-building and conclude in Section \ref{sec:conclusion}.

%% file: tuning.tex
To motivate the structure of the Twin Turtle in general, and the Twin Maximally Natural Turtle in particular, we begin by reviewing aspects of fine-tuning in the original Twin Higgs model before introducing the simplest turtle extension and illustrating its parametric advantages. This necessarily entails a careful accounting of the full sensitivity of electroweak symmetry breaking to UV physics, expanding upon previous treatments of fine-tuning in Twin Higgs models.

\subsection{Twin Higgs Tuning}
\label{subsec:review}
To understand tuning in the original Twin Higgs model, we begin with the most general linear sigma potential for a pair of complex scalars $h_a, h_b$, each of which transforms as a doublet under respective $SU(2)_{a}$ and $SU(2)_b$ gauge groups, which in turn are related by a discrete $\mathbb{Z}_2$ exchange symmetry:
\begin{equation}
\label{eq:twinpot}
V =  \lambda (|h_a|^2 + |h_b|^2)^2 + m_h^2 (|h_a|^2 + |h_b|^2) + \kappa (|h_a|^4 + |h_b|^4) + \rho |h_a|^4 + \mu^2 |h_a|^2.
\end{equation}
When $\kappa / \lambda, \rho / \lambda, \mu^2/m_h^2 \ll 1$, this theory has an approximate $SU(4)$ global symmetry. For appropriate signs in the potential, both $h_a$ and $h_b$ acquire vacuum expectation values, $\langle |h_a|^2 \rangle + \langle |h_b|^2 \rangle = f^2$, leading to a pseudo-goldstone of the spontaneously broken $SU(4)$ that is identified with the SM-like Higgs. In a full Twin Higgs model, we identify $SU(2)_{a}$ and $SU(2)_b$ with the weak gauge groups of two copies of the Standard Model, and extend the discrete symmetry to exchange SM$_{a}$ with SM$_{b}$. We moreover identify SM$_{a}$ as ``our'' copy of the Standard Model, and $SU(2)_a$ as our corresponding weak gauge group.

It's convenient to study the properties of the SM-like Higgs $\phi$ in the pNGB limit, in which case it can be related to $h_a, h_b$ and the scale of $SU(4)$ breaking $f$ via
\begin{equation}
h_a = f\sin \frac{\phi}{\sqrt{2}f}, ~~h_b=f\cos\frac{\phi}{\sqrt{2}f}.
\end{equation}
Inserting this parameterization into Eq.~(\ref{eq:twinpot}) leads to expressions for the electroweak symmetry breaking \textit{vev} $v$ and SM-like Higgs mass $m_{SM}^2$ of the form
\begin{eqnarray}
\label{eq:twinapprox}
2 v^2 =f^2\frac{2\kappa}{2\kappa +\rho} -\frac{\mu^2}{2\kappa+\rho}, \\ \nonumber
m^2_{SM} = 4 v^2 (2\kappa +\rho) (1-\frac{v^2}{f^2}) \,.
\end{eqnarray}
At this point it is straightforward to determine the tuning of the electroweak scale with respect to underlying parameters. Treating the Twin Higgs model as an effective theory up to some cutoff scale $\Lambda$, we would ultimately like to determine the sensitivity of the Higgs \textit{vev} $v$ to the cutoff, allowing for the possibility that the scale $f$ is intermediate between the two. 
 
It is common to approximate the overall tuning of $v$ by factorizing it in terms of the sensitivity of $v$ to the scale $f$, $\Delta_{v/f}$, and the sensitivity of $f$ scale to the cutoff scale $\Delta_{f/\Lambda}$, $\Delta^{\twin}_{v/\Lambda} = \Delta_{v/f} \times \Delta^{\twin}_{f/\Lambda}$ (e.g. \cite{Craig:2013fga}). Here the factorization of tuning is valid insofar as the weak scale only depends implicitly on the cutoff through an intermediate scale, as is the case for the Twin Higgs potential in Eq.~(\ref{eq:twinpot}) at tree level. This factorization can be violated by a variety of effects, including both logarithmic dependence on the cutoff that arises from radiative corrections to quartic couplings and quadratic dependence on the cutoff from radiative corrections via hard-breaking quartics. In what follows, the (in)validity of factorization will often be important in understanding the fine tuning of Twin Turtles.

Both factorized and non-factorized contributions to the tuning of the weak scale can be accounted for succinctly by treating the fine-tuning in terms of the total derivative of the Higgs \textit{vev} with respect to the cutoff, which in the current case takes the schematic form
\begin{equation}
\Delta^{\twin}_{v/\Lambda} \equiv \frac{d \log v^2}{d \log \Lambda^2} = \frac{\partial \log v^2}{\partial \log f^2} \frac{d \log f^2}{d \log \Lambda^2} + \frac{\partial \log v^2}{\partial \log \Lambda^2} .
\end{equation}
The first term reproduces the usual factorized result, while the second term accommodates direct sensitivity of $v$ to $\Lambda$ due to violations of factorization. Let us consider them each in turn. \\

The first term involves direct sensitivity of $v$ to the scale $f$, and sensitivity of $f$ to the cutoff(s) $\Lambda$. For the Twin Higgs, the sensitivity of $f$ to the cutoff scale is simply that of a complex scalar with various couplings to SM-like fields. The two largest couplings are those associated with the top yukawa $y_t$ in each sector and the $SU(4)$-symmetric quartic $\lambda$, and so it is convenient to distinguish the cutoff $\Lambda_t$ associated with the top sector from the cutoff $\Lambda_\rho$ associated with the scalar sector. (Here we will not distinguish between the cutoffs associated with the $SU(4)$-symmetric quartic $\lambda$ and the $SU(4)$- and $\mathbb{Z}_2$-breaking quartic $\rho$, though of course the two may be quite distinct in a UV completion.) Then the tuning of the scale $f$ with respect to these cutoffs is simply
\begin{eqnarray} \label{eq:tuningf}
 \frac{\partial \log f^2}{\partial \log \Lambda_t^2} &=&   \frac{3}{32 \pi^2} \frac{y_t^2}{\lambda} \frac{\Lambda_t^2}{f^2}, \\
 \frac{\partial \log f^2}{\partial \log \Lambda_\rho^2} &=&  - \frac{5}{32\pi^2} \frac{\Lambda_{\rho}^2}{f^2}   \, .
\end{eqnarray} 
The tuning of $v$ with respect to $f$ depends on the nature of the parameters breaking the $\mathbb{Z}_2$ symmetry. If  the $\mathbb{Z}_2$ symmetry is only broken softly via the $\mu^2$ term in Eq.~\eqref{eq:twinpot}, such that $\rho = 0$, we have
\begin{equation}
\Delta^{\soft}_{v/f}  = \left. \frac{\partial \log v^2}{\partial \log f^2} \right |_{\rho = 0} =  \frac{f^2}{2 v^2},
\end{equation}
corresponding to the familiar tuning of typical Twin Higgs models. But as was illustrated in~\cite{Katz:2016wtw}, if the $\mathbb{Z}_2$ symmetry is broken instead through the hard-breaking quartic $\rho$, such that $\mu^2=0$, we have
\begin{equation}
\label{eq:tuninghard}
\Delta^{\hard}_{v/f} = \left. \frac{\partial \log v^2}{\partial \log f^2} \right |_{\rho \neq 0} = \left. \frac{2\kappa}{2\kappa+\rho} \times  \frac{\partial \log v^2}{\partial \log f^2} \right |_{\rho = 0}   =\frac{2\kappa}{2\kappa+\rho} \frac{f^2}{2v^2} \, .
\end{equation}
When $\kappa\ll\rho$, $\Delta_{v/f}$ can be improved significantly, in principle erasing tuning associated with the separation between $v$ and $f$. However, $\frac{2\kappa}{2\kappa+\rho}$ cannot be arbitrarily small thanks to two effects. The first is the Higgs mass constraint, i.e. the contribution of $2 \kappa + \rho$ to the SM-like Higgs quartic, whose value is known. The second is the irreducible contribution to the quartic $\kappa$ from fermion loops. The main radiative correction from the top loop to $\kappa$ is given by
\begin{equation}
\delta\kappa = \frac{3 y^4_t}{16\pi^2} \log{\frac{\Lambda^2_t}{y^2_t f^2}},
\end{equation}
which on its own is not far from the SM-like Higgs quartic. These two considerations bound the improvement in fine-tuning from the introduction of a hard-breaking quartic $\rho$ to 
\begin{equation}\label{eq:hardlimit}
1 \gtrsim \frac{\Delta^{\hard}_{v/f}}{\Delta^{\soft}_{v/f} } \gtrsim \frac{3}{8 \pi^2} \frac{y_t^4}{\lambda_{SM}} \log \frac{\Lambda_t^2}{y_t^2 f^2} \gtrsim \frac{1}{\rm few} \, .
\end{equation}

Now we turn to the second contribution to the tuning of $v$ coming from the direct sensitivity of $v$ to the cutoff. When the $\mathbb{Z}_2$ symmetry is only broken softly via the $\mu^2$ term, this contribution vanishes, and we obtain the familiar factorized result. However, additional hard breaking of the $\mathbb{Z}_2$ symmetry via the quartic $\rho$ restores quadratic sensitivity of the SM-like Higgs mass to the cutoff $\Lambda_\rho$ at one loop,
\begin{equation}
\delta \mu^2 =  \epsilon_\pm \frac{3 \rho}{16\pi^2}\Lambda_\rho^2 \, ,
\label{eq:epm}
\end{equation}
where $\epsilon_\pm = \pm 1$ depends on the UV completion. 
This amounts to a modest violation of the factorization assumption, made tolerable only by the smallness of $\rho \lesssim \lambda_{SM}$, and
\begin{equation}
\frac{\partial \log v^2}{\partial \log \Lambda_\rho^2} \approx - \frac{3\epsilon_\pm}{32 \pi^2} \frac{ \rho}{2 \kappa + \rho} \frac{\Lambda_\rho^2}{v^2}
\end{equation}
where we neglect contributions coming from logarithmic dependence on the cutoffs.

Putting everything together, the total tuning in the case of soft $\mathbb{Z}_2$ breaking, up to logarithmic cutoff dependence, is thus
\begin{equation}
\Delta^{\twin,\soft}_{v/\Lambda} =  \frac{1}{64 \pi^2} \left(\frac{3y^2_t}{\lambda} \frac{\Lambda^2_t}{v^2} - 5 \frac{\Lambda^2_\rho}{v^2} \right)
\end{equation}
while the tuning in the presence of hard $\mathbb{Z}_2$ breaking is
\begin{equation}
\Delta^{\twin,\hard}_{v/\Lambda} =   \frac{1}{64 \pi^2} \frac{2 \kappa}{2 \kappa + \rho} \left(\frac{3y^2_t}{\lambda} \frac{\Lambda^2_t}{v^2} - \left( 5 + 3 \epsilon_\pm \frac{ \rho}{\kappa} \right) \frac{\Lambda^2_\rho}{v^2}  \right) \ .
\end{equation}

\subsection{A Twin Turtle}

\begin{figure}
\centering
\includegraphics[scale=0.35]{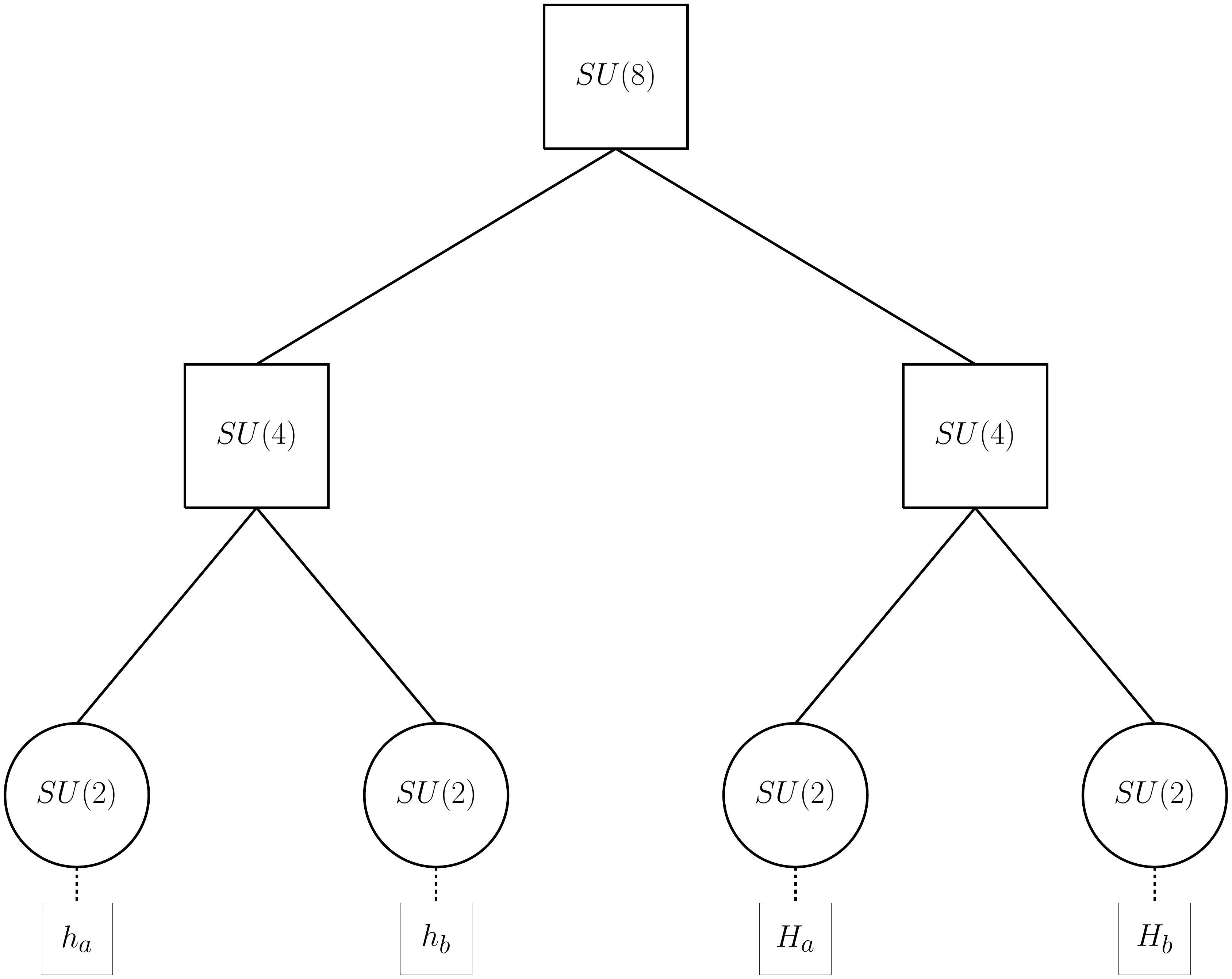}
\caption{ The symmetry structure of the Twin Turtle model pursued in this work, in which a softly-broken Twin Higgs model is UV completed by yet another Twin Higgs model with certain hard-breaking quartics. Only the $SU(2)$ groups are gauged; the $SU(4)$ and $SU(8)$ are approximate global symmetries. }
\label{fig:setup}
\end{figure}

It is apparent in Eq.s~\eqref{eq:tuninghard}-\eqref{eq:hardlimit} that a significant improvement in tuning would be possible if hard breaking were responsible for separating scales unrelated to the SM-like Higgs. This naturally suggests constructing a Twin Turtle in which a Twin Higgs model with a softly-broken discrete symmetry is UV completed by a Twin Higgs structure involving hard breaking of a discrete symmetry. This setup is illustrated in Fig.~\ref{fig:setup}. As there are no mass constraints on the radial mode of the new twin structure, the constraints on the size of the hard breaking in the new upstairs twin (owing to the SM-like Higgs mass) are removed and we can potentially reach a larger improvement in the total fine tuning.

To this end, we envision a model with four identical copies of the Standard Model. Each copy contains a complex scalar, which we will respectively label $h_a, h_b, H_a,$ and $H_b$, transforming as a doublet under its own $SU(2)$ gauge symmetry. As with the Twin Higgs, we will associate the Higgs doublet $h_a$, with corresponding weak group $SU(2)_a$, with ``our'' copy of the Standard Model.

 The four identical copies of the Standard Model in isolation nominally enjoy an $S_4$ symmetry acting on the field labels, but this is generally broken by turning on couplings between Higgs scalars from different copies. Rather, with the symmetry breaking structure of turtles in mind, it is convenient to instead organize the theory around the following pattern of discrete symmetries relating the Higgs scalars (and their corresponding copies of the Standard Model): The doublets $h_a$ and $h_b$ are related by a discrete $\mathbb{Z}_2$ symmetry, as are $H_a$ and $H_b$, leading to two approximate $SU(4)$ global symmetries acting on $h \equiv (h_a, h_b)$ and $H \equiv (H_a, H_b)$. An exchange symmetry relating these two sectors then gives rise to an approximate $SU(8)$ global symmetry of the theory. In this respect it is useful to think of the symmetry of the Higgs scalars as $[SU(2) \times SU(2) \times \mathbb{Z}_2] \times [SU(2) \times SU(2) \times \mathbb{Z}_2] \times \mathbb{Z}_2$. The potential for the Higgs scalars will break these discrete symmetries in a hierarchical fashion. Needless to say, while this is a natural extension of the Twin Higgs model to a turtle structure, it is not the unique way of constructing a Twin Turtle, and other interesting configurations of symmetry embeddings are possible.

In what follows, we will often refer to the approximately $SU(8)$-symmetric sector as the ``upstairs theory'', and the approximate $SU(4)$-symmetric sector containing the Standard Model $SU(2)_a$ as the ``downstairs theory.''  The approximate $SU(8)$ global symmetry is broken at a scale $F$, and the $SU(4)$ global symmetry containing the Standard Model $SU(2)_a$ is broken at a scale $f < F$. 

Starting from the approximate $SU(8)$ global symmetry, the relevant terms in the potential for this model are given by 
\begin{eqnarray}
\label{eq:fullpot}
V & \supset & \bar{\lambda} (|h|^2 + |H|^2)^2 + m^2 (|h|^2+|H|^2) + \bar{\kappa} (|h|^4+|H|^4) + \bar{\rho} |h|^4 + m_h^{\prime 2} |h|^2 \\
&+& \rho^{\prime\prime} (|h_a|^2+|H|^2)^2 +  \kappa'' (|h_a|^4+|h_b|^4+|H_a|^4+|H_b|^4) + \kappa^\prime (|h_a|^4+|h_b|^4) + \rho^\prime |h_a|^4 + \mu^{\prime 2} |h_a|^2. \nonumber
\end{eqnarray}

Let us briefly summarize the role of each coupling:
\begin{itemize}

\item $m^2$ and $\bar{\lambda}$ respect the full $SU(8)$ global symmetry of the model and determine the scale $F$ of $SU(8)$ symmetry breaking.

\item $\bar{\kappa}$ only breaks the approximate $SU(8)$, while preserving the $SU(4) \times SU(4) \times \mathbb{Z}_2$ symmetry. This term is crucial for developing additional scales below $F$, and in particular determines the scale $f$. 

\item $\bar{\rho}$ and $m_h'^2$ break the upstairs $\mathbb{Z}_2$ (i.e.~the $\mathbb{Z}_2$ in $SU(4) \times SU(4) \times \mathbb{Z}_2$), while preserving the individual $SU(4)$ symmetries (and the $SU(2) \times SU(2) \times \mathbb{Z}_2$ symmetries therein).

\item $\rho^{\prime\prime}$ respects our $SU(2)\times SU(2)$, while breaking our $\mathbb{Z}_2$ (i.e. the $\mathbb{Z}_2$ in the $SU(2) \times SU(2) \times \mathbb{Z}_2$ containing $SU(2)_a$) and hence also our $SU(4)$.

\item $\kappa^{\prime\prime}$ breaks both $SU(4)$ symmetries but respects the $[SU(2) \times SU(2) \times \mathbb{Z}_2] \times [SU(2) \times SU(2) \times \mathbb{Z}_2]  \times \mathbb{Z}_2$ symmetry. This is crucial for developing \textit{vev}s below the scales of $SU(4)$ symmetry breaking on both sides.

\item $\kappa^\prime$ breaks both the upstairs $\mathbb{Z}_2$ symmetry and the approximate $SU(4)$ containing $SU(2)_a$, preserving the $SU(2) \times SU(2) \times \mathbb{Z}_2$ containing $SU(2)_a$. 

\item $\rho^\prime$, $\mu^{\prime 2}$ break all the global symmetries in our model. $\rho^\prime$ is analogous to the hard quartic responsible for $\mathbb{Z}_2$ breaking in the original Twin Higgs model, while $\mu^{\prime 2}$ is analogous to the soft-breaking mass term. 

\end{itemize}

The couplings $\bar{\lambda}$, $\bar{\kappa}$, and $\bar{\rho}$ evidently play the same role as $\lambda$, $\kappa$, and $\rho$, respectively, in Eq.~\eqref{eq:twinpot}. Neglecting possible additional $\mathbb{Z}_2$ breaking between $H_a$ and $H_b$ in the $SU(4)$ not containing $SU(2)_a$ (as it will not affect any of our conclusions), Eq.~\eqref{eq:fullpot} is the most general renormalizable potential we can write for our structure. In what follows, we will assume the $\rho^{\prime\prime}$ coupling is negligible compared to other symmetry-breaking quartics. This will be justified by the UV completions presented in Sec.~\ref{sec:topdown}, for which this quartic only arises as a two-loop effect.

Working in the pNGB limit at each step, the vacuum expectation values and mass spectrum can be approximated as 
\begin{eqnarray}
\label{eq:feq}
&2 f^2& \sim \frac{2\bar{\kappa}}{2\bar{\kappa} +\bar{\rho}} F^2 - \frac{m^{\prime 2}_h}{2\bar{\kappa}+\bar{\rho}},~~
2 v^2 \sim f^2-\frac{\mu^{\prime 2}}{2\kappa_1}, \\ \nonumber
&m^2_1& \sim 4 \bar{\lambda} F^2,~~
m^2_2 \sim 4 (2\bar{\kappa} + \bar{\rho}) f^2,~~
m^2_3 \sim 4 \kappa'' \frac{F^2-f^2}{2}, \\ \nonumber
&m^2_{SM} &\sim 4 v^2 (2\kappa_1) (1-\frac{v^2}{f^2}),
\end{eqnarray}
where we have defined $\kappa_1 = \kappa^{\prime\prime} +\kappa^\prime$. 

The tuning of the weak scale with respect to the cutoffs $\Lambda_t$ and $\Lambda_\rho$ of the top and Higgs sectors takes the general form
\begin{eqnarray}
\Delta_{v/\Lambda}^{\tmnt} = \frac{\partial \log v^2}{\partial \log f^2} \frac{d \log f^2}{d \log \Lambda^2} + \frac{\partial \log v^2}{\partial \log F^2} \frac{d \log F^2}{d \log \Lambda^2} + \frac{\partial \log v^2}{\partial \log \Lambda^2} 
\end{eqnarray}
where we are taking care to accommodate violations of factorization. For the pattern of hard and soft breaking taken here, we have $\partial v / \partial F \approx 0$ and $\partial v / \partial \Lambda \approx 0$ up to logarithmic dependence. Since the separation of $v$ from $f$ arises only through soft breaking of a $\mathbb{Z}_2$, we have the usual
\begin{eqnarray}
\label{eq:vftuning}
\frac{\partial \log v^2}{\partial \log f^2} = \frac{f^2}{2 v^2}.
\end{eqnarray}
The nontrivial contribution to the tuning lies in the dependence of the scale $f$ on the cutoff, where (in analogy with the above discussion of fine-tuning in the Twin Higgs with hard $\mathbb{Z}_2$ breaking)
\begin{eqnarray}
\frac{d \log f^2}{d \log \Lambda^2}  &=& \frac{\partial \log f^2}{\partial \log F^2} \frac{d \log F^2}{d \log \Lambda^2} + \frac{\partial \log f^2}{\partial \log \Lambda^2}  \\
& \approx & \left[ \frac{2 \bar \kappa}{2 \bar \kappa + \bar \rho} \frac{F^2}{2 f^2}\right] \left[ \frac{1}{32 \pi^2} \left( \frac{3 y_t^2}{\bar \lambda} \frac{\Lambda_t^2}{F^2} - 9 \frac{\Lambda_\rho^2}{F^2} \right) \right] - \frac{1}{32 \pi^2} \frac{5 \epsilon_\pm \, \bar \rho}{2 \bar \kappa + \bar \rho} \frac{\Lambda_\rho^2}{f^2} .
\end{eqnarray}
Thus the tuning of the weak scale with respect to the cutoff in this Twin Turtle is, up to logarithmic corrections,
\begin{eqnarray}
\Delta_{v/\Lambda}^{\tmnt} \approx \frac{1}{64 \pi^2} \frac{\bar \kappa}{2 \bar \kappa + \bar \rho} \left( \frac{3 y_t^2}{\bar \lambda} \frac{\Lambda_t^2}{v^2} - \left( 9 + 5 \epsilon_\pm \frac{\bar \rho}{\bar \kappa} \right) \frac{\Lambda_\rho^2}{v^2} \right).
\end{eqnarray}

\subsection{Comparison with SM and Twin Higgs}
\label{subsec:FT}

What have we gained? As a reminder, the tuning in the SM and a Twin Higgs model (with either soft or hard breaking of the $\mathbb{Z}_2$ symmetry, keeping track of the factorization violation in the latter case) with respect to cutoffs $\Lambda_t$ and $\Lambda_\rho$ takes the form
\begin{eqnarray}
\label{eq:twinsmft}
\Delta^{\twin,\soft}_{v/\Lambda} &\approx&  \frac{1}{64 \pi^2} \left(\frac{3y^2_t}{\lambda} \frac{\Lambda^2_t}{v^2} - 5 \frac{\Lambda^2_\rho}{v^2} \right) , \\ \nonumber 
\Delta^{\twin,\hard}_{v/\Lambda} &\approx&   \frac{1}{64 \pi^2} \frac{2 \kappa}{2 \kappa + \rho} \left(\frac{3y^2_t}{\lambda} \frac{\Lambda^2_t}{v^2} - \left( 5 + 3 \epsilon_\pm \frac{\rho}{\kappa} \right) \frac{\Lambda^2_\rho}{v^2}  \right),
 \\ \nonumber 
\Delta^{\sm}_{v/\Lambda} &\approx& \frac{1}{32\pi^2}\left(\frac{3 y^2_t}{\lambda_{\rm SM}}\frac{\Lambda^2_t}{v^2} - 3 \frac{\Lambda^2_\rho}{v^2} \right) \, .
\end{eqnarray}
where $\lambda$ is the approximately $SU(4)$-symmetric quartic in the Twin Higgs and $\lambda_{\rm SM}$ is the SM Higgs quartic.

Let us first consider the sensitivity to $\Lambda_t$. In many respects, this is the most important cutoff from the perspective of searches for new colored states associated with naturalness of the weak scale, in that it corresponds to the mass scale of colored top partners in a UV completion.\footnote{The top quarks in the additional copies of the Standard Model serve as the top partners in the low-energy effective theory, but are themselves not charged under our copy QCD.} The fine tuning improvement in the TMNT over a Twin Higgs model or the SM will be
\begin{eqnarray}
\label{eq:tcomp}
\frac{\Delta^{\tmnt}_{v/\Lambda_t}}{\Delta^{\twin,\soft}_{v/\Lambda_t}} &=& \frac{\lambda}{\bar{\lambda}} \frac{\bar{\kappa}}{2\bar{\kappa} +\bar{\rho}}, \\
\frac{\Delta^{\tmnt}_{v/\Lambda_t}}{\Delta^{\twin,\hard}_{v/\Lambda_t}} &=& \frac{\lambda}{\bar{\lambda}} \frac{\bar{\kappa}}{2 \kappa} \frac{2 \kappa + \rho}{2\bar{\kappa} +\bar{\rho}} \approx \frac{1}{2} \frac{\lambda}{\bar \lambda} \frac{\lambda_{\rm SM}}{2 \bar \kappa + \bar \rho}, \\
 \frac{\Delta^{\tmnt}_{v/\Lambda_t}}{\Delta^{\sm}_{v/\Lambda_t}} &=& \frac{\lambda_{\rm SM}}{2\bar{\lambda}}\frac{ \bar{\kappa}}{2\bar{\kappa} +\bar{\rho}}.
\end{eqnarray}
The tuning improvement relative to the Standard Model is straightforward. The tuning improvement over the Twin Higgs depends on the Twin Higgs quartic $\lambda$ relative to the TMNT quartic $\bar \lambda$. Since the TMNT involves more states that contribute to the running of $\bar{\lambda}$, we expect $\bar{\lambda}<\lambda$ assuming a common Landau pole scale, which undermines the gain in tuning. For example, in a SUSY UV completion, to have the same Landau poles, $\bar{\lambda} \simeq \frac{1}{2}\lambda$.  To compensate, we require $\frac{ \bar{\kappa}}{2\bar{\kappa} +\bar{\rho}}< \frac{1}{2}$ in order for the tuning of the TMNT with respect to $\Lambda_t$ to improve over soft-breaking Twin Higgs models, which is straightforward. To see an improvement relative to hard-breaking Twin Higgs models, we require $\frac{\bar \kappa}{2 \kappa} \frac{2 \kappa + \rho}{2\bar{\kappa} +\bar{\rho}} \approx \frac{1}{2} \frac{\lambda_{\rm SM}}{2 \bar \kappa + \bar \rho}< \frac{1}{2}$, which is also quite feasible.

Crucially, there are two structural features that allow both $2 \bar \kappa + \bar \rho \ll 2 \kappa + \rho$ and $\bar \kappa \ll \bar \rho$, apart from any constraints arising from the need to reproduce the Standard Model Higgs quartic. First, Yukawa interactions do not correct $\bar{\kappa}$ at one loop, in contrast to the $\kappa$ parameter in Twin Higgs models that is generated from the top yukawa. Second, the dominant radiative correction to $\bar{\kappa}$ is proportional to $\bar{\rho}$, 
\begin{equation}
\delta \bar{\kappa}=\frac{5 \bar{\rho}\bar{\lambda}}{32\pi^2}  \log \frac{\Lambda^2_\rho}{m^2_2} 
\label{eq:dbarkappa}
\end{equation}
which permits $\bar{\kappa}$ to be small, and $\bar \kappa \ll \bar \rho$, without additional fine-tuning.

Next we consider the sensitivity to $\Lambda_\rho$.  The tuning improvement over Twin Higgs models and the SM with respect to $\Lambda_\rho$ is
\begin{eqnarray}
\label{eq:rhocomp}
\frac{\Delta^{\tmnt}_{v/\Lambda_\rho}}{\Delta^{\twin,\soft}_{v/\Lambda_\rho}} &=& \frac{1}{5} \frac{9\bar{\kappa} + 5 \epsilon_\pm \bar{\rho}}{2\bar{\kappa} +\bar{\rho}}, \\
\frac{\Delta^{\tmnt}_{v/\Lambda_\rho}}{\Delta^{\twin,\hard}_{v/\Lambda_\rho}} &=& \frac{9 \bar \kappa + 5 \epsilon_\pm \bar \rho}{2 \bar \kappa + \bar \rho} \frac{2 \kappa + \rho}{10 \kappa + 6 \epsilon_\pm \rho},
\\
\frac{\Delta^{\tmnt}_{v/\Lambda_\rho}}{\Delta^{\sm}_{v/\Lambda_\rho}} &=& \frac{1}{6} \frac{9\bar{\kappa}+5 \epsilon_\pm \bar{\rho}}{2\bar{\kappa} +\bar{\rho}}.
\end{eqnarray}
For $\bar{\rho} \gg \bar{\kappa}$ -- a sensible limit, as noted above -- the sensitivity of the electroweak scale to $\Lambda_\rho$ is comparable to the SM and softly-broken Twin Higgs, and potentially somewhat better than the hard-breaking Twin Higgs. In order to avoid severe fine-tuning, new electroweak states associated with the Higgs sector should appear around 2~TeV. 

One interesting feature is apparent from these comparisons: while improving the sensitivity to the cutoff of the top sector relative to the Standard Model, the Twin Higgs model does not improve the sensitivity with respect to new electroweak states associated with the cutoff of the Higgs sector. Moreover, the price paid by the TMNT model in order to push the cutoff of the top sector to even higher scales is the introduction of {\it two} additional radial modes around the TeV range. This trade-off is even more apparent when working in terms of the fine-tuning of the Higgs mass instead of the electroweak scale, which we discuss in the next subsection.

\subsection{Fine Tuning {\it Lausannois}}
\label{subsec:FT2}

We conclude our discussion of fine-tuning by re-phrasing the results of the previous subsection in the language of \cite{Contino:2017moj}, which frames fine-tuning in Twin Higgs models intuitively in terms of leading corrections to the Higgs doublet mass parameter. While the tuning of the doublet mass parameter is equivalent to the tuning of the \textit{vev} discussed above (as the quartic is not tuned), this provides a clear setting for understanding the key features of tuning in turtled models, as well as straightforward comparison with the authoritative discussion of Twin Higgs tuning in \cite{Contino:2017moj}. Here we assume that vacuum alignment is such that the Higgs mass parameter is identifiable with the electroweak doublet $h_a$ in both the Twin Higgs and TMNT models. 

In the Standard Model, the dominant radiative corrections to the Higgs doublet mass parameter are
\begin{equation}
\delta m^2_{\sm} = -\frac{3y^2_t}{16\pi^2}\Lambda^2_t +\frac{3 \lambda_{SM}}{16\pi^2} \Lambda^2_\rho.
\label{eq:smtuning}
\end{equation}
In the Twin Higgs model with soft $\mathbb{Z}_2$ breaking, the analogous expression is
\begin{equation}
\delta m^2_{\sm}= \frac{\kappa}{\lambda}(m^2_{h} -\frac{3y^2_t}{16\pi^2}\Lambda^2_t + \frac{5 \lambda}{16\pi^2} \Lambda^2_\rho) + \mu^2,
\label{eq:twintuning}
\end{equation}
where $2\kappa \sim \lambda_{\rm SM}$. As we have seen, compared to the Standard Model, the sensitivity to the colored top partner scale is reduced by the factor of $\frac{\kappa}{\lambda}$ and the fine tuning with respect to the scalar cutoff is roughly the same as that of the SM. Finally, the price of improved tuning in the top sector is the introduction of a radial mode of mass $\sim |m_h|$.

Extending the Twin Higgs to include the hard $\mathbb{Z}_2$ breaking term $\rho |h_a|^4$, we have
\begin{equation}
\delta m^2_{\sm}= \frac{\kappa}{\lambda}(m^2_{h} -\frac{3y^2_t}{16\pi^2}\Lambda^2_t + \frac{5 \lambda}{16\pi^2} \Lambda^2_\rho) + \frac{3 \epsilon_\pm \rho}{16\pi^2} \Lambda^2_\rho+ \mu^2,
\label{eq:twintuning2}
\end{equation}
with $(2\kappa+\rho)$ fixed to be roughly $\lambda_{\rm SM}$. One would like to have as small a value of $\kappa$ as possible to reduce the fine tuning. However, as we have discussed above, the radiative correction to $\kappa$ from top loops almost saturate the SM quartic $\lambda_{\rm SM}$, limiting the improvement of fine tuning. 

Turning now to the TMNT potential in Eq.~\eqref{eq:fullpot} (without the $\rho^\prime$ and $\rho^{\prime\prime}$ quartics for simplicity), we have
\begin{eqnarray}
\delta m^2_{\sm} &=& \frac{\kappa_1}{2\bar{\kappa} + \bar{\rho}}\bigg[ \frac{\bar{\kappa}}{\bar{\lambda}} \bigg(m^2 - \frac{3y^2_t}{16\pi^2}\Lambda^2_t + \frac{9\bar{\lambda}}{16\pi^2}\Lambda^2_\rho\bigg) + \frac{5 \epsilon_\pm \bar{\rho}}{16\pi^2} \Lambda^2_\rho \bigg] + \mu^{\prime 2}\\ \nonumber
&=&  \frac{\kappa_1}{\bar{\lambda}}\frac{\bar{\kappa}}{2\bar{\kappa} + \bar{\rho}}\bigg(m^2- \frac{3y^2_t}{16\pi^2}\Lambda^2_t +\frac{9\bar{\lambda}}{16\pi^2}\Lambda^2_\rho\bigg) + \frac{\epsilon_\pm \bar{\rho}}{2\bar{\kappa}+\bar{\rho}}\frac{5\kappa_1 \Lambda^2_\rho}{16\pi^2} +\mu^{\prime 2},
\label{eq:ustuning}
\end{eqnarray}
where $2\kappa_1\sim\lambda_{\rm SM}$. Insofar as $2\bar{\kappa}+\bar{\rho}$ is not constrained by known quartics, the sensitivity to $\Lambda_t$ can be improved significantly by taking $\bar{\rho}\gg\bar{\kappa}$. As discussed earlier, this limit is readily attainable since radiative corrections to $\bar \kappa$ from Yukawa couplings are absent, and $\bar \kappa$ may be separated from $\bar \rho$ by as much as a loop factor. This limit has little impact on the sensitivity to the cutoff associated with Higgs quartics, which is again comparable to the Standard Model. Finally, there are now {\it two} radial modes appearing in the theory, both of which lie at or below the scale $|m|$.

\begin{figure}
\centering
\includegraphics[width=0.5\textwidth]{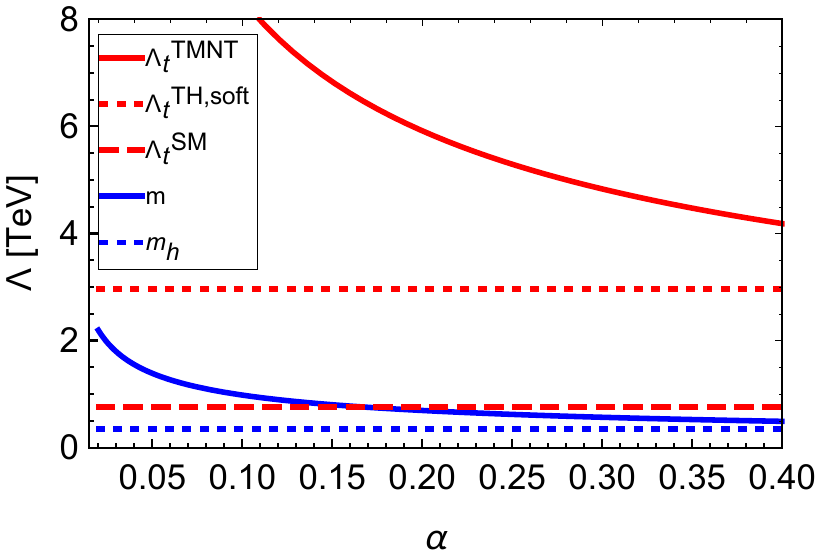}
\caption{Natural values for the top sector cutoff $\Lambda_t$ and radial modes in the TMNT model, the softly-broken Twin Higgs model, and the Standard Model as a function of $\alpha \equiv \frac{\bar{\kappa}}{2\bar{\kappa}+\bar{\rho}}$. The solid red line, dotted red line, and dashed red line denote the respective top sector cutoff scale in each model corresponding to no tuning of the Higgs mass. The natural values for the Higgs quartic cutoff are comparable in each model -- $\sim \unit[2]{TeV}$ -- and are not shown here. The solid blue line and dotted blue line denote the tree-level mass parameters $m$ and $m_h$ of the TMNT and softly-broken Twin Higgs, respectively, which are proxies for the scale of new radial modes. Here we have taken $\kappa_1=\kappa=0.065, \bar{\lambda}=0.8,\lambda=1.0, y_t=0.85$. }
\label{fig:cutoff}
\end{figure}

To better illustrate the sensitivity of the electroweak scale to different UV scales in each of these models, in \fig{cutoff} we show the natural values of the top sector cutoff $\Lambda_t$ (corresponding to an un-tuned Higgs mass) and the mass scale of new radial modes ($m_h$ in the Twin Higgs, $m$ in the TMNT) as a function of the parameter $\alpha \equiv \frac{\bar{\kappa}}{2\bar{\kappa}+\bar{\rho}}$.  For smaller values of $\alpha$, the cutoff $\Lambda_t$ associated with new colored states in the top sector can be raised well above the corresponding scale in Twin Higgs model and the Standard Model.  In contrast, the tree-level symmetric scales setting the masses of the radial modes are required to be lighter than the scale associated with colored top partners, emphasizing the tradeoff between colored top partners and new states in the Higgs sector. It is interesting to notice that the cutoff $\Lambda_\rho$ associated with scalar loops is roughly of the same magnitude in all three models, which indicates that we should expect new electroweak states at the few-TeV level irrespective of other considerations. 

Given the success of improving fine-tuning with respect to $\Lambda_t$ by introducing an additional radial mode, one might be tempted to further improve fine-tuning by stacking additional turtles with hard $\mathbb{Z}_2$-breaking quartics on top of each other. However, as discussed earlier, the hard-breaking quartics reintroduce sensitivity to the Higgs quartic cutoff at one loop, so no improvement in the scale of new electroweak-charged states can be expected. Moreover, the one-loop logarithmic dependence of the Higgs quartic on the cutoff of the top sector ultimately limits the extent to which the scale of colored top partners can be pushed off. Taken together, these effects tend to saturate the improvement in fine-tuning with one turtle unless extensive model-building gymnastics (relative to those already in play) are employed. As such, we will limit ourselves to one additional level of symmetry breaking in the TMNT.  Having demonstrated the parametric improvement in fine-tuning, we now turn to a more detailed exploration of the model.

%% file: bottomup.tex
In this section we explore the TMNT model illustrated in Fig.~\ref{fig:setup} in further detail as an effective theory with cutoffs $\Lambda_t$ and $\Lambda_\rho$, with an eye towards radiative correlations between the parameters of the potential and the qualitative phenomenology.  In its entirety, the model consists of four copies of the Standard Model, each with a Higgs doublet ($h_a, h_b, H_a, H_b$, respectively). The matter content, gauge groups, gauge couplings, and yukawa couplings of each copy are identical. The only couplings between these copies of the Standard Model are via the Higgs sector, for which the potential relating the different Higgs fields (subject to simplifications and assumptions discussed in Sec. \ref{sec:tuning}) is

\begin{eqnarray}
\label{eq:fullpot2}
V & \supset & \bar{\lambda} (|h|^2 + |H|^2)^2 + m^2 (|h|^2+|H|^2) + \bar{\kappa} (|h|^4+|H|^4) + \bar{\rho} |h|^4 + m^{\prime 2}_h |h|^2 \\
&+& \rho^{\prime\prime} (|h_a|^2+|H|^2)^2 +  \kappa'' (|h_a|^4+|h_b|^4+|H_a|^4+|H_b|^4) + \kappa^\prime (|h_a|^4+|h_b|^4) + \rho^\prime |h_a|^4 + \mu^{\prime 2} |h_a|^2. \nonumber
\end{eqnarray}
Radiative corrections correlate the parameters in the potential and give rise to both quadratic and logarithmic dependence on the cutoffs $\Lambda_t$ and $\Lambda_\rho$. In particular, the dominant one-loop corrections take the form
\begin{eqnarray}
\label{eq:loops}
\delta m^2 &=& -\frac{3 y^2_t}{16\pi^2} \Lambda^2_t + \frac{9\bar{\lambda}}{16\pi^2} \Lambda_\rho^2, \\ \nonumber
\delta \bar{\kappa} &=& \frac{5 \bar{\rho}\bar{\lambda}}{32\pi^2}  \log \frac{\Lambda^2_\rho}{m^2_2}, \\ \nonumber
\delta \kappa^{\prime\prime}&=& \frac{3 y^4_t}{16\pi^2}\log\frac{\Lambda^2_t}{m^2_{T_a}},\\ \nonumber
\delta \kappa^{\prime} &=& \frac{3 y^4_t}{16\pi^2}\log\frac{m^2_{T_a}}{m^2_{t_b}} ,\\ \nonumber
\delta \rho^{\prime} &=& \frac{3 y^4_t}{16\pi^2}\log\frac{m^2_{t_b}}{m^2_{t_a}} ,\\ \nonumber
\delta m^{\prime 2}_h &=& \epsilon_\pm \frac{5 \bar{\rho}}{16\pi^2} \Lambda^2_\rho ,
\end{eqnarray}
where $m_{t_b}$ ($m_{T_a}$) is the mass of the top twin in the $h_b$ ($H_a$) side, and we have used the fact that the mass $m_{T_b}$ of the top twin in the $H_b$ side is equal to $m_{T_a}$ as the $\mathbb{Z}_2$ within $H$ is respected. We have neglected the log term correction for $\delta m^{\prime 2}_h$ assuming that the quadratically divergent contribution dominates. In addition, we have neglected radiative contributions coming from electroweak gauge couplings, which are generally subdominant.

\subsection{Fine Tuning}

Having discussed the parametrics of fine-tuning in Sec. \ref{sec:tuning}, we now perform a more detailed numerical study that takes into account logarithmic sensitivity to the cutoff and accommodates the observed value of the Higgs mass. As before, we concentrate on tuning with respect to dimensionful parameters, as tuning with respect to dimensionless parameters is always subleading in the regime of interest. 

For simplicity, we illustrate the numerical tuning with respect to a generic benchmark point, corresponding to the parameters in \tab{epsilon}. To explore the tuning we consider varying $\Lambda_t$ and the tree-level value of $\bar \rho$, using $\mu^\prime$ to then set the electroweak $vev$. As we will justify in the next section, it is possible to have only some of the couplings generated at tree-level and others induced purely radiatively. Given the tree level value of each parameter in  \tab{epsilon}, we include the radiative corrections in Eq.~\eqref{eq:loops}.
\begin{table}
\begin{center}
\begin{tabular}{|c|c|c|c|c|c|c|c|c|}
\hline 
$m~[\mathrm{TeV}]$  &$\Lambda_\rho~[\mathrm{TeV}]$ & $\bar{\lambda}$ & $\bar{\kappa}$ & $\kappa''$ & $\kappa'$ & $\rho'$&$m^2_h$ & $\rho^{\prime\prime}$ \\ 
\hline 
1.8 &3.0&0.8 & 0.0(0.15)& 0.0 & 0.005 & 0.0&0.0 &0.0  \\ 
\hline 
\end{tabular} 
\caption{
Tree-level parameters used for our numerical study of fine-tuning with respect $\Lambda_t$ and tree-level $\bar{\rho}$. Numbers in parentheses are for $\epsilon_{+}$. 
}
\label{tab:epsilon}
\end{center}
\end{table}

With these corrections included,  we investigate the fine-tuning of the scalar sector in the $(\Lambda_t-\bar{\rho})$ plane. At this stage we will adopt simplified notation and drop the ``TMNT'' superscripts on the fine-tuning measure $\Delta$, as henceforth all fine-tunings are calculated for the TMNT model.
In order to capture a parametrization-independent result, we add the contribution of all the parameters in quadrature and define the total tuning as
\begin{equation}
\Delta_v = \sqrt{\max \lbrace 1, \sum_x \left(\frac{x}{v^2} \frac{\partial v^2}{\partial x} \right)^2 \rbrace },
\label{eq:FTquad}
\end{equation}
where $x = \lbrace \mu'^2,m^2, m_h^{\prime 2}, \Lambda_t^2, \Lambda_\rho^2\rbrace$ runs over all of the dimensionful inputs. This will give us a conservative measure of the tuning of the EW scale and separate the irreducible tuning of the structure in Fig.~\ref{fig:setup} from the tuning due to any specific underlying model. We will turn to tuning calculations in UV-complete supersymmetric models in the next section.

As we have already discussed in the previous section, it is often convenient to discuss the overall tuning of the electroweak scale by factorizing it into the product of tunings capturing the sensitivity of $v$ to the scale $f$, and then $f$ to the cutoff scale $\Lambda$, although this fails to capture general logarithmic cutoff depence as well as additional quadratic sensitivity of $v$ to $\Lambda$ in the presence of hard breaking. To illustrate the extent to which the factorization approximation captures the fine-tuning, we also define a factorized tuning measure of the form 
\begin{equation}
\Delta_{v} \sim \Delta_{v/f} \times  \Delta_{f/\Lambda} = \sqrt{\max \lbrace 1, \sum_{x_v} \left(\frac{x_v}{v^2} \frac{\partial v^2}{d x_v} \right)^2 \rbrace }  \times \sqrt{1+\sum_{x_f} \left(\frac{x_f}{f^2} \frac{\partial f^2}{\partial x_f} \right)^2 },
\label{eq:FTfactor}
\end{equation}
where $x_v = \lbrace		\mu'^2\rbrace$ 
, while $x_f= \lbrace m^2, m_h^{\prime 2},  \Lambda_t^2, \Lambda_\rho^2	\rbrace$. This will allow us to illustrate the extent to which the factorization approximation captures the leading contributions to fine-tuning.

\begin{figure}
\begin{center}
\includegraphics[scale=0.65]{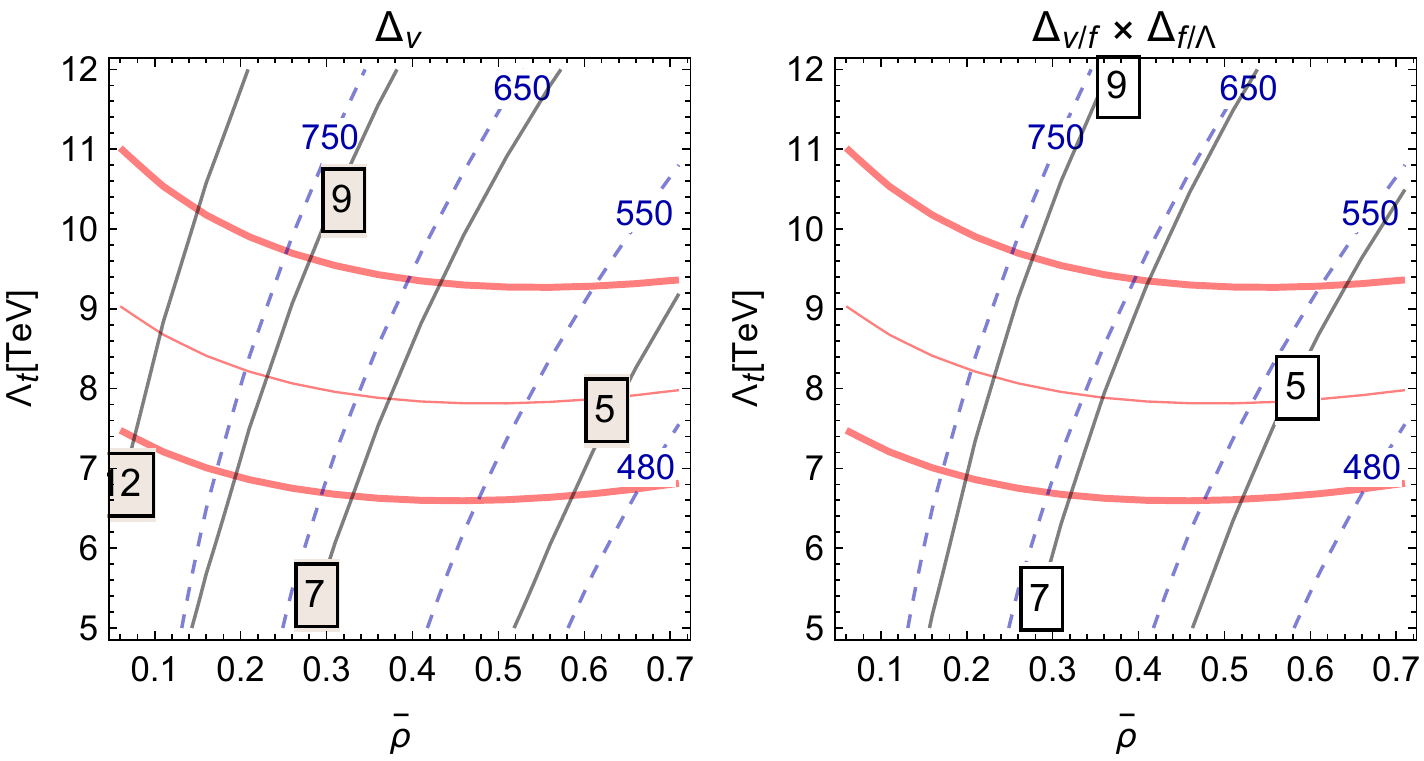}
\caption{Solid contours denote the total tuning $\Delta_v$ of $v$ with respect to all parameters added in quadrature (left) and the factorized approximation $\Delta_{v/f}\times \Delta_f$ (right) as a function of $\bar \rho$ and $\Lambda_t$ for $\epsilon_{+}$. The parameters fixed numerically are reported in \tab{epsilon}. The red band denotes the Higgs mass window (122 GeV, 128 GeV). The dashed line denotes the scale $f$ in unit of GeV.}
\label{fig:numFTvtot_ep}
\end{center}
\end{figure}

For the case of $\epsilon_+$, the radiative correction $\delta m^{\prime 2}_h$  from the $\bar{\rho}$ term tends to stabilise the vacuum of $h$ around zero. A relatively large tree-level $\bar{\kappa}$ is needed to compensate this and induce spontaneous breaking in the $h$ sector. We thus choose relatively large $\bar{\kappa}\sim 0.15$ to study the total fine tuning in this case. The total tuning of the electroweak $vev$ $\Delta_v$ and the factorization approximation $\Delta_{v} \times\Delta_{v/f}$ are shown in \fig{numFTvtot_ep}. Also shown are the contours of the scale $f$.
For the chosen parameters, the factorization approximation is in good agreement with the total tuning. As can be seen, the total tuning $\Delta_v$ parallels contours of increasing $f$, reflecting the fact that the dominant contribution to fine-tuning is coming from the soft breaking separating $v$ from $f$. 

\begin{figure}
\begin{center}
\includegraphics[scale=0.65]{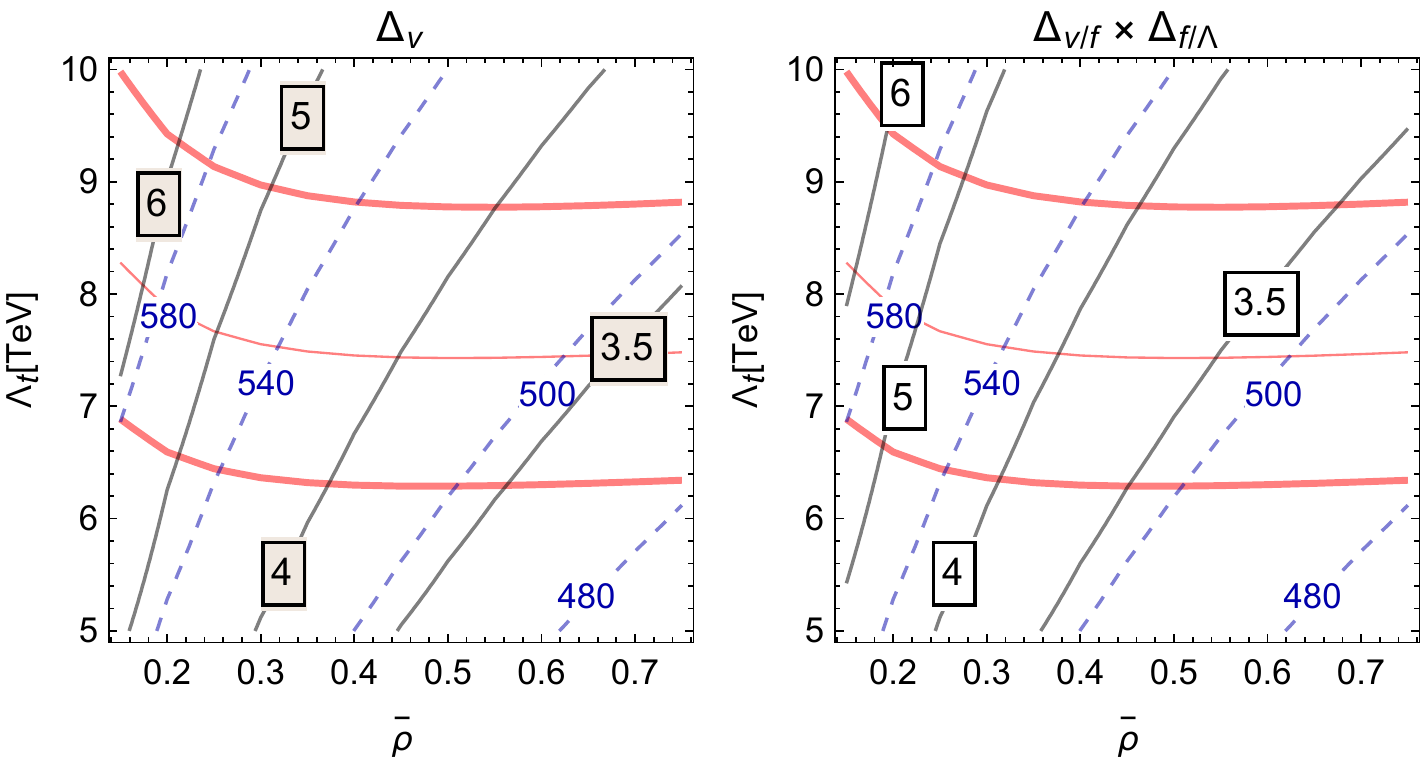}
\caption{Solid contours denote the total tuning $\Delta_v$ of $v$ with respect to all parameters added in quadrature (left) and the factorized approximation $\Delta_{v/f}\times \Delta_f$ (right) as a function of $\bar \rho$ and $\Lambda_t$ for $\epsilon_{-}$. The parameters fixed numerically are reported in \tab{epsilon}. The red band denotes the Higgs mass window (122 GeV, 128 GeV). The dashed line denotes the scale $f$ in unit of GeV.}
\label{fig:numFTvtot}
\end{center}
\end{figure}

The case of $\epsilon_{-}$ turns out to be somewhat different. As there are no requirements on the tree-level $\bar{\kappa}$, we choose $\bar{\kappa}=0$ for the sake of reducing fine tuning. As before, the total tuning of the electroweak $vev$ $\Delta_v$ and the factorization approximation $\Delta_{v} \times\Delta_{v/f}$ are shown in \fig{numFTvtot}, demonstrating good agreement between the two. For this parameter space, $F\sim\unit[1.5]{TeV}$ and slightly increases with $\Lambda_t$. Since $\bar{\kappa}$ and $m^{\prime 2}_h$ are purely radiatively generated from $\bar{\rho}$, while $m^{\prime 2}_h$ is quadratically sensitive to $\Lambda^2_\rho$, the scale $f$ is primarily determined by $\delta m^{\prime 2}_h$ and doesn't change much as a function of $\bar{\rho}$.

\begin{figure}
\begin{center}
\centering
\includegraphics[scale=0.60]{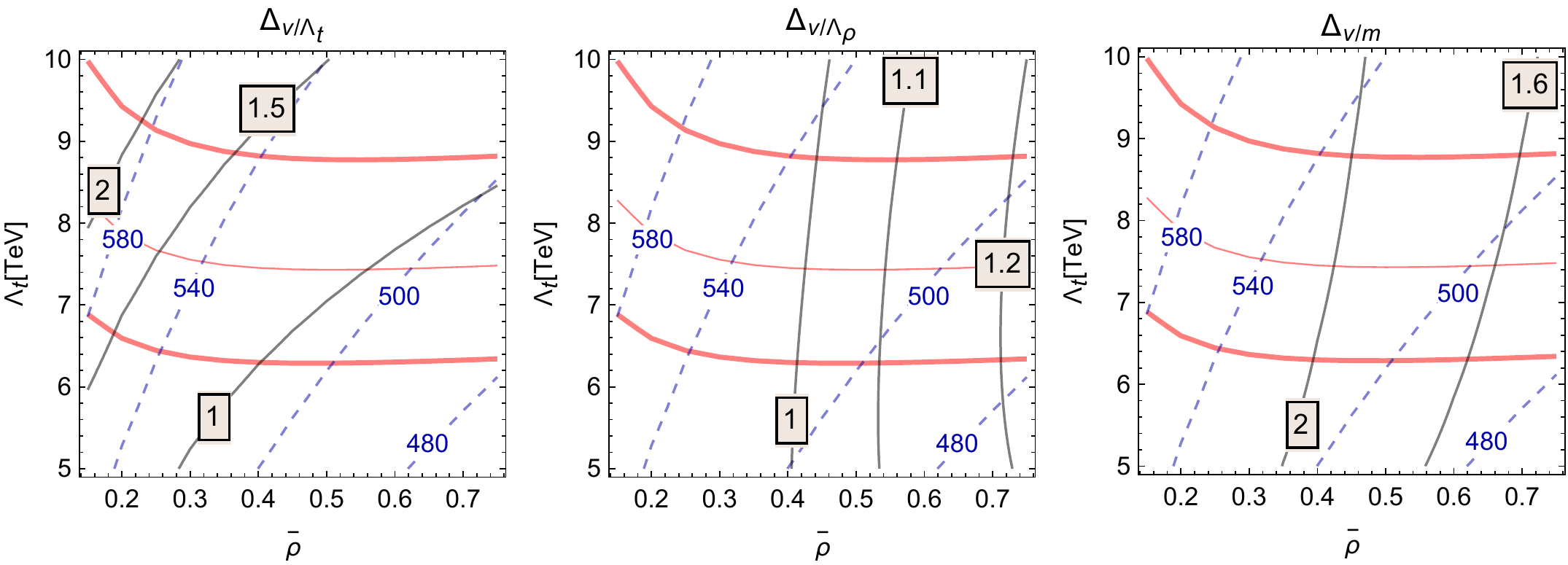}
\caption{The total tuning of $v$ with respect to the mass parameters $m^2, \Lambda^2_t$ and $\Lambda^2_\rho$ as a function of $\bar \rho$ and $\Lambda_t$. The red band denotes the Higgs mass window (122 GeV, 128 GeV), while the dashed line indicates the value of the scale $f$ in unit of GeV.}
\label{fig:numFTv}
\end{center}
\end{figure}

We next turn to consider tuning of the electroweak $vev$ with respect to the three high energy scales $m^2, \Lambda^2_t$ and $\Lambda^2_\rho$ in the $\bar{\rho}-\Lambda_t$ plane, illustrated in \fig{numFTv}. 
It is apparent from the tuning with respect to $\Lambda_t$ that the colored top partner can be as high as $\unit[9]{TeV}$ without introducing much tuning, provided a sufficiently large $\bar{\rho}$. Also noteworthy is the fact that the tuning with respect to $\Lambda_t$ is not quadratically decreasing with $\Lambda_t$ at low values of the cutoff, which is quite distinct from the vanilla Twin Higgs model. This is because the reduced quadratic sensitivity to $\Lambda_t$ implies that the primary sensitivity of the weak scale to $\Lambda_t$ in this regime comes from the one loop logarithmic dependence in Eq.~\eqref{eq:loops}. Although the factorization approximation is not shown, it remains a good proxy for the total fine-tuning in this range of parameters despite factorization violations from $\delta\kappa^\prime$ and $\delta\kappa^{\prime\prime}$.

As the TMNT structure doesn't improve the quadratic sensitivity to $\Lambda^2_\rho$, the tuning with respect to $\Lambda^2_\rho$ is still dominated by the quadratic dependence and requires $\Lambda_\rho$ to be much smaller than $\Lambda_t$ for $\mathcal{O}(1)$ tuning. The natural scale of the tree-level symmetric mass parameter $m$ must also be much smaller than $\Lambda_t$ since the separation of the electroweak scale from this scale is set by $\frac{\kappa_1}{\bar{\lambda}}\frac{\delta \bar{\kappa}}{2\delta \bar{\kappa}+\bar{\rho}}$. While the tuning with respect to $m$ may in principle be further reduced relative to the values in \fig{numFTv} by decreasing $m$, this is ultimately limited by the need to maintain sufficiently large $m$ (and hence $f$) to satisfy precision Higgs coupling constraints. 

Our discussion of tuning in this section neglects the origin of the terms in the general potential from Eq.~\eqref{eq:fullpot}. In this respect, they represent the intrinsic and irreducible tuning of the electroweak scale in our setup, which may be further increased by detailed UV completions. 

\subsection{Phenomenology}

\begin{figure}[ht]
\begin{center}
\includegraphics[scale=0.5]{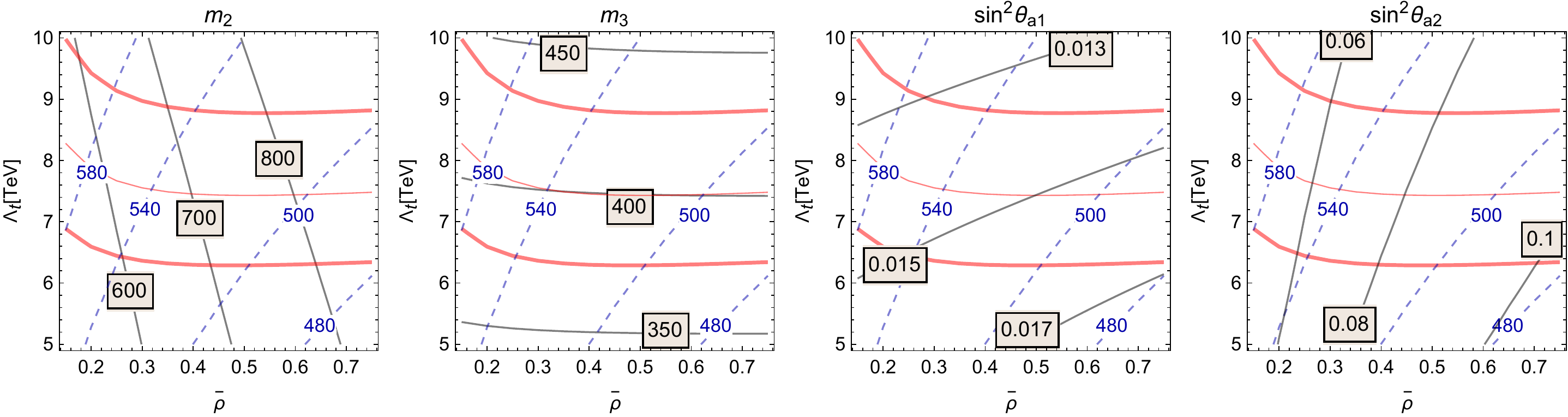}
\caption{From left to right: The mass $m_2$ of the approximate radial mode of $SU(4) \supset SU(2)_a$; the mass $m_3$ of the scalar associated with the other $SU(4)$; the mixing angle $\sin^2 \theta_{a1}$ between the SM doublet $h_a$ and the radial mode of $SU(8)$; and the mixing angle $\sin^2 \theta_{a2}$ between the SM doublet $h_a$ and the approximate radial mode of $SU(4) \supset SU(2)_a$. In each frame, the red band denotes the Higgs mass window (122 GeV, 128 GeV), while the dashed line indicates the scale of $f$ in unit of GeV.  The numerical values used in this plot are those  reported in \tab{epsilon} for $\epsilon_{-}$.}
\label{fig:nummassmix}
\end{center}
\end{figure}

We next turn to a brief exploration of the infrared phenomenology of the TMNT model. As noted earlier, the improvement in cutoff sensitivity comes at the price of additional Higgs scalars. In addition to the SM-like Higgs boson, there are three new CP-even neutral scalars $h_1$, $h_2$ and $h_3$, corresponding to the radial mode associated with breaking the approximate $SU(8)$ and the two other uneaten goldstones of $SU(8) \rightarrow SU(7)$ breaking. Of these two additional goldstones, $h_2$ can be roughly thought of as the radial mode of $SU(4) \supset SU(2)_a$, while $h_3$ is a scalar associated with the other $SU(4)$ sector. The masses of these scalars are roughly given by $m_1$, $m_2$ and $m_3$, respectively in Eq.~\eqref{eq:feq}. For $\epsilon_{-}$ with the parameters shown in \tab{epsilon}, $m_1$ is around $\unit[2.5]{TeV}$. The masses $m_2$, $m_3$ of the two lighter modes and the two largest mixing angles $\theta_{a1}$ and $\theta_{a2}$ relating the SM doublet $h_a$ to the $SU(4) \supset SU(2)_a$ and $SU(8)$ radial modes are illustrated in \fig{nummassmix}. As a consequence of mixing, the SM-like Higgs couplings are modified relative to the Standard Model by an amount
\begin{equation}
g_{hii} \sim g_{h_{SM} ii}\bigg(1- \frac{1}{2}(\sin^2 \theta_{a1}+\sin^2 \theta_{a2})\bigg) \sim g_{h_{SM} ii}(1- \frac{v^2}{2f^2})
\end{equation}
where the second approximation, much as in the vanilla Twin Higgs, is numerically apparent in \fig{nummassmix}.
This leads to the dominant bound on the scale $f$ due to constraints on Higgs coupling deviations. Given Higgs coupling measurements consistent with Standard Model predictions at the $\mathcal{O}(10\%)$ level \cite{Craig:2013fga,Buttazzo:2015bka}, the scale $f$ is constrained to lie above $\sim 500$ GeV. By increasing $m^2$, one can increase the scale $f$ without changing the sensitivity with respect to the cutoffs $\Lambda_t$ and $\Lambda_\rho$. While smaller $\bar{\lambda}$ or larger $\bar{\kappa}$ would also increase the scale $f$, the fine tuning with respect to the cutoff $\Lambda_t$ would correspondingly increase. 

With the colored top partners associated with the scale $\Lambda_t$ well out of reach of the LHC, the additional Higgs scalars are the most promising avenue for discovering the Twin Turtle at the LHC. All three additional neutral Higgs scalars inherit their couplings to the Standard Model via mixing with the SM doublet $h_a$. Since the $SU(8)$ radial mode $h_1$ is very heavy and contains only a small admixture of $h_a$, the prospects for probing $h_1$ are somewhat remote, but in some sense constitutes the smoking gun signal of the Twin Turtle.

The lighter mode $h_2$ plays a role similar to the radial model in the vanilla Twin Higgs model if we identify $2\bar{\kappa}+\bar{\rho}$ with the vanilla Twin Higgs $SU(4)$ symmetric quartic coupling. In contrast to the vanilla Twin Higgs, however, for a given SM-like Higgs coupling deviation, the mixing between $h_2$ and $h_a$ is reduced by roughly a factor of $f^2/F^2$ owing to the mixing of $h_1$ and $h_a$. This indicates that the coupling of $h_2$ to SM particles is reduced by the same factor, compared to the vanilla Twin Higgs radial mode. Thus, given a certain SM-like Higgs coupling deviation, the production cross section of $h_2$ at LHC is smaller. As the coupling reduction is universal, the branching ratios of $h_2$ to different Standard Model channels is not changed. This highlights the value of making precise rate and branching ratio measurements of any heavy Higgs scalars produced in the future, as it allows differentiation between a vanilla Twin Higgs scenario and a Twin Turtle.

In addition to the radial modes of $SU(8)$ and $SU(4) \supset SU(2)_a$, there is the third scalar $h_3$ associated with the other $SU(4)$ sector. If $\kappa^{\prime\prime}$ is loop-generated, this could be the lightest of the new scalar modes, as shown in \fig{nummassmix}. However, its mixing with $h_a$ is highly suppressed because of the degeneracy between $H_a$ and $H_b$, and its production rate at the LHC is correspondingly small.

%% file: topdownF.tex
We begin with a supersymmetric UV completion of the TMNT model that uses $F$-terms to generate the leading quartics.  As always, the introduction of supersymmetry also mandates the doubling of the Higgs spectrum, with each $SU(2)$ doublet now promoted to a pair of doublets, which we denote with $u$ and $d$ subscripts and refer to as up-type and down-type doublets, respectively. The $\mathbb{Z}_2$ symmetries acting on the doublets of the low-energy TMNT model are naturally extended to act on both the up-type and down-type doublets in the UV completion. 

For the $F$-term supersymmetric model, we need to introduce additional singlets to generate different quartics. To generate a sizable $SU(8)$ quartic, one singlet $S_0$ is introduced to couple to the $SU(8)$-symmetric combination of Higgs doublets. In order to ensure that this singlet generates an $SU(8)$-symmetric quartic, we need only require that its superpotential coupling respect the exchange symmetry relating $h_u \equiv (h^a_u, h^b_u)$ and $h_d \equiv (h_{d}^a, h_{d}^b)$ to $H_u \equiv (H^a_u, H^b_u)$ and  $H_d \equiv (H^a_d, H^b_d)$. This is sufficient to generate an $SU(8)$-invariant quartic, in analogy with supersymmetric completions of the vanilla Twin Higgs \cite{Chang:2006ra, Falkowski:2006qq, Craig:2013fga}.

The $SU(8)$-symmetric superpotential and the soft masses are
\begin{eqnarray}
W_{SU(8)} &=& (\mu +\lambda_{s0} S_0) (H_u H_d + h_u h_d) + \frac{\mu_{s0}}{2} S_0^2, \\
V_{SU(8)} &=& m^2_{u} (|H_u|^2+|h_u|^2) +m^2_{d} (|H_d|^2+|h_d|^2) \\ \nonumber
&-& b(H_u H_d + h_u h_d + h.c.) + m_{s0}^2 S_0^2 .
\end{eqnarray}
Assuming that the singlet soft mass $m_{s0}$ is much larger than the supersymmetric mass $\mu_{s0}$, we can integrate out $S_0$ non-supersymmetrically to obtain the following potential for the Higgs degrees of freedom:
\begin{eqnarray}
V^{eff}_{SU(8)}& =& (m^2_{u}+\mu^2) (|H_u|^2+|h_u|^2) +(m^2_{d} +\mu^2) (|H_d|^2+|h_d|^2)\\ \nonumber
 &-& b(H_u H_d + h_u h_d+h.c.)+\lambda^2_{s0} |H_u H_d + h_u h_d|^2
\end{eqnarray}
where we are suppressing corrections subleading in $\mu_{s0}^2 / m_{s0}^2$.

In order to generate quartics respecting the two $SU(4) \subset SU(8)$ symmetries, we introduce an additional two singlet superfields $S_h$ and $S_H$. In contrast to $S_0$, these superfields are related by the same $\mathbb{Z}_2$ that exchanges $H\leftrightarrow h$. If the potential for $S_h$ and $S_H$ fully respected the same exchange symmetry, then this would suffice to generate the $\bar \kappa$ quartic but not the $\bar \rho$ quartic. To generate $\bar \rho$, we allow for a soft breaking of the exchange symmetry relating $S_h$ and $S_H$. Upon integrating out $S_h$ and $S_H$, this will feed down into an effective quartic that violates the $\mathbb{Z}_2$ symmetry relating the two $SU(4)$s. Since $\mathbb{Z}_2$ is broken in the UV lagrangian only by soft masses, the one loop radiative corrections to the dimensionless couplings still preserve the $\mathbb{Z}_2$.

The superpotential and the soft masses for this level of the theory are
\begin{eqnarray}
W_{\cancel{SU(8)}}&=& \lambda_{sh} S_h h_u h_d+ \lambda_{sH} S_H H_u H_d + \frac{\mu_{sh}}{2} S^2_h + \frac{\mu_{sH}}{2} S^2_H ,\\
V_{\cancel{SU(8)}} &=& m^2_{sh} |S_h|^2 + m^2_{sH} |S_H|^2 ,
\end{eqnarray}
where $\lambda_{sh} =\lambda_{sH}$ and $\mu_{sh} =\mu_{sH}$ at tree level because of $\mathbb{Z}_2$ symmetry. A further restriction is necessary in order to forbid additional terms in the superpotential or soft potential involving both $S_h$ and $S_H$. The simplest possibility is to impose independent PQ symmetries on both the $h$ and $H$ sectors. 

Integrating out $S_h$ and $S_H$, we obtain the $SU(8)$-breaking potential for the Higgs sector,
\begin{eqnarray}
V^{eff}_{\cancel{SU(8)}}& =&\frac{\lambda^2_{sh}  m^2_{sh}}{\mu^2_{sh}+m^2_{sh}} |h_u h_d|^2 + \frac{\lambda^2_{sH} m^2_{sH}}{\mu^2_{sH}+m^2_{sH}}|H_u H_d|^2.
\end{eqnarray}
 
The spectrum of the theory now includes a variety of additional Higgs scalars implied by the additional Higgs doublets. However, these additional scalars can be lifted in the decoupling limit of the 2HDM in each copy of the MSSM, so that the spectrum of light states is precisely that of the TMNT model. Assuming equal $t_\beta$ for each sectors, we can match the full potential to the non-supersymmetric potential $V$ of the low-energy TMNT by rotating to the Higgs basis
\begin{eqnarray}
h_u = h s_{\beta}, \, \, h_d= h c_{\beta} , \\ \nonumber
H_u = H s_{\beta}, \, \, H_d= H c_{\beta},
\end{eqnarray}
with 
\begin{equation}
t^2_{\beta} =\frac{m^2_d + \mu^2}{m^2_u + \mu^2},~~
\end{equation}
The matching between UV parameters and the low-energy TMNT parameters is simply
\begin{eqnarray}
\label{eq:matchup}
m^2& =&  (m^2_{u}+\mu^2) s^2_\beta +  (m^2_{d}+\mu^2) c^2_\beta - b s_ {2\beta}, \\ \nonumber
\bar{\lambda} &=&\frac{\lambda^2_{s0}}{4} s^2_{2\beta}, \\ \nonumber
\bar{\kappa} &=&\frac{\lambda^2_{sH} m^2_{sH}}{\mu^2_{sH}+ m^2_{sH}}\frac{s^2_{2\beta}}{4}, \\ \nonumber
\bar{\rho}&=&\left(\frac{\lambda^2_{sh} m^2_{sh}}{\mu^2_{sh} + m^2_{sh}} -\frac{\lambda^2_{sH} m^2_{sH}}{\mu^2_{sH} + m^2_{sH}}\right)\frac{s^2_{2\beta}}{4}.
\end{eqnarray}
We see that in the region where $m^2_{sH}\ll \mu^2_{sH}$ and $m^2_{sh}\gg \mu^2_{sh}$, the parametric ratio $\bar{\kappa} \ll \bar{\rho}$ is realized. Notice that $m^2_{sH}$ cannot be arbitrarily small, as the soft mass terms $m^2_u$ and $m^2_d$ contribute to $m^2_{sH}$ at one loop level, setting the natural size for $m^2_{sH}$. 

Electroweak gauge couplings and Yukawa interactions, while preserving $\mathbb{Z}_2$ symmetries, will contribute to $SU(8)$ breaking quartics at one loop. These contributions are
\begin{eqnarray}
V_{\cancel{SU(8)}}& =& \frac{g^2_1+g^2_2}{8} \bigg[(|h^a_u|^2-|h^a_d|^2)^2 +(|h^b_u|^2-|h^b_d|^2)^2 \\ \nonumber
&+&(|H^a_u|^2-|H^a_d|^2)^2 +(|H^b_u|^2-|H^b_d|^2)^2\bigg]\\ \nonumber
&+&\frac{3 y^4}{16\pi^2}\bigg[ \log\left(\frac{m^2_{\tilde{t}_a}}{m^2_{t_a}}\right)|h^a_u|^4+  \log\left(\frac{m^2_{\tilde{t}_b}}{m^2_{t_b}}\right)|h^b_u|^4+ \log\left(\frac{m^2_{\tilde{T}_a}}{m^2_{T_a}}\right)|H^a_u|^4+ \log\left(\frac{m^2_{\tilde{T}_b}}{m^2_{T_b}}\right)|H^b_u|^4\bigg] ,
\end{eqnarray}
where $m_{\tilde{t}_a}, m_{\tilde{t}_b}, m_{\tilde{T}_a}$ and $m_{\tilde{T}_b}$ are the stop masses\footnote{Common LH and RH stop masses in each sector are assumed.} in the $h_a$, $h_b$, $H_a$ and $H_b$ sectors, respectively.  At tree level, we have  $m_{\tilde{t}_a}= m_{\tilde{t}_b}=m_{\tilde{T}_a} = m_{\tilde{T}_b} =m_{\tilde{t}}$.
These radiative corrections contribute to the effective potential for the TMNT model via the matching
\begin{eqnarray}
\label{eq:matchdown}
\kappa^{\prime\prime} &=& \frac{g^2_{ew}}{8}c^2_{2\beta} +\frac{3 y_t^4}{16\pi^2}\log\left(\frac{m^2_{\tilde{t}}}{m^2_{T_a}}\right),\\ \nonumber
\kappa^\prime &=& \frac{3 y_t^4}{16\pi^2} \log\left(\frac{m^2_{T_a}}{m^2_{t_b}}\right),
\end{eqnarray}
where $y_t = y \sin \left(\beta \right)$, $g^2_{ew} = g^2_1+g^2_2$ is the electroweak gauge coupling in the SM, and we have used the relation $m^2_{T_a} = m^2_{T_b}$ as the $\mathbb{Z}_2$ within $H$ is respected.

The most important radiative corrections to the $SU(8)$ symmetric mass are from the stop loop and singlet $S_0$ loop,
\begin{eqnarray}
\delta m^2_u &\sim& -\frac{3 y^2}{16\pi^2}m^2_{\tilde{t}}\log\left(\frac{\Lambda_{mess}^2}{m^2_{\tilde{t}}}\right) - \frac{\lambda^2_{s0}}{16\pi^2}m^2_{s0}\log\left(\frac{\Lambda^2_{mess}}{m^2_{s0}}\right), \\ \nonumber
\delta m^2_d&\sim&- \frac{\lambda^2_{s0}}{16\pi^2}m^2_{s0}\log\left(\frac{\Lambda^2_{mess}}{m^2_{s0}}\right),
\end{eqnarray}
where $\Lambda_{mess}$ is the messenger scale. In what follows we will typically take $\Lambda_{mess}=10 m_{\tilde{t}}$.

The soft mass terms $m^2_{sh}$ and $m^2_{sH}$ will also feed into $m^{\prime 2}_h$ at one loop via, for example,
\begin{equation}
\label{eq:matchsoft}
m^{\prime 2}_h =-\frac{\lambda^2_{sh} m^2_{sh}}{16\pi^2}\log\left(\frac{\Lambda^2_{mess}}{m^2_{sh}+\mu^2_{sh}}\right).
\end{equation}
Radiative corrections from $m^2_u$ and $m^2_d$ to $m^{\prime 2}_h$ are finite and neglected in this section.
The sign of $m_h^{\prime 2}$ is negative as we have $m^2_{sh} \gg m^2_{sH}$ and positive soft mass terms $m^2_{sh}$, $m^2_{sH}$ in order not to induce non-zero vacuum expectation values in the singlet sectors. Though will not be studied here, we note that corrections from the singlet soft mass terms could be improved by introducing additional singlets as in \cite{Lu:2013cta}.  The $m^{\prime 2}_h$ term further contributes to $t_\beta$ on the $h$ side as
\begin{equation}
t^2_{\beta_h} = \frac{m^2_d + \mu^2 + m^{\prime 2}_h}{m^2_u + \mu^2 + m^{\prime 2}_h}.
\end{equation}
As the correction only makes $t^2_{\beta_h}$ slightly different from $t^2_{\beta}$, we will neglect it in what follows.

As the $\mathbb{Z}_2$ within $h$ has to be softly broken to generate the misalignment between $f$ and electroweak $vev$, we introduce the following soft mass terms
\begin{equation}
V_{\cancel{SU(4)}} = \delta m^2_u |h^a_u|^2 +\delta m^2_d |h^a_d|^2 +\delta b (h^a_u h^a_d +h.c.)
\end{equation} 
which amount to soft breaking of the $\mathbb{Z}_2$ symmetry relating $h_a$ and $h_b$. Matching to the TMNT potential $V$ gives $\mu^{\prime 2} = (\delta m^2_u s^2_\beta + \delta m^2_d c^2_\beta)$, with $\delta b=0$. This misalignment also leads to a modest contribution to the downstairs quartic, 
\begin{equation}
\label{eq:rhop}
\Delta V^\prime = \frac{3 y_t^4}{16\pi^2} \log\left(\frac{m^2_{t_b}}{m^2_{t_a}}\right)|h_a|^4= \rho^\prime |h_a|^4  .
\end{equation}

\subsubsection{Constraints \& Fine Tuning}

We are now in a position to consider the tuning of the electroweak scale in the context of a UV completion. Before studying the tuning in detail, it is necessary to consider the viable range of couplings in the UV completion. In particular, in order to obtain sizeable quartics in the infrared TMNT model, the UV couplings ($\lambda_{s0}, \lambda_{sh}, \lambda_{sH}$) should also be sizeable. However, these couplings cannot be too large without inducing a Landau pole close to the scale of matching, compromising the validity of the UV completion.  

The one-loop beta functions for the couplings $\lambda_{s0}, \lambda_{sh}$ and $\lambda_{sH}$ are
\begin{eqnarray}
16\pi^2\frac{d}{dt} \lambda_{s0} &=& \lambda_{s0}(10\lambda^2_{s0} + 6\lambda^2_{sh}+3 y^2_t -g^2_Y -3 g^2_2),\\ \nonumber
16\pi^2\frac{d}{dt} \lambda_{sh} &= &\lambda_{sh} (6\lambda^2_{sh} + 6\lambda^2_{s0}+3 y^2_t -g^2_Y -3 g^2_2),\\ \nonumber
16\pi^2\frac{d}{dt} \lambda_{sH}&= &\lambda_{sH} (6\lambda^2_{sH} + 6\lambda^2_{s0}+3 y^2_t -g^2_Y -3 g^2_2).
\label{eq:running}
\end{eqnarray}
If we require that neither coupling hits a Landau pole below $\unit[100]{TeV}$, this implies that the couplings in the infrared ($\unit[5]{TeV}$, for concreteness) must satisfy
\begin{eqnarray}
\lambda_{s0} < 1.3,~~\lambda_{sh} =\lambda_{sH} < 1.9 .
\end{eqnarray}
For comparison, in the $F$-term supersymmetric completion of the vanilla Twin Higgs model, the singlet coupling is constrained to satisfy the weaker condition $\lambda_{s0}< 2.1$ to avoid a Landau pole beneath 100 TeV. This exemplifies the point made in Sec. \ref{sec:tuning} regarding the relative sizes of symmetric quartics in the TMNT and vanilla Twin Higgs.

\begin{figure}
\centering
\includegraphics[scale=0.75]{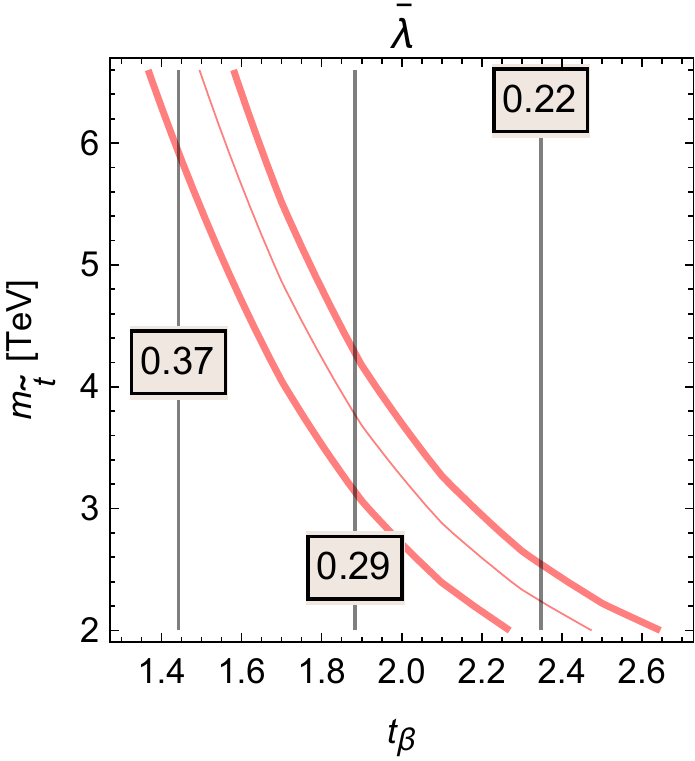}
\includegraphics[scale=0.75]{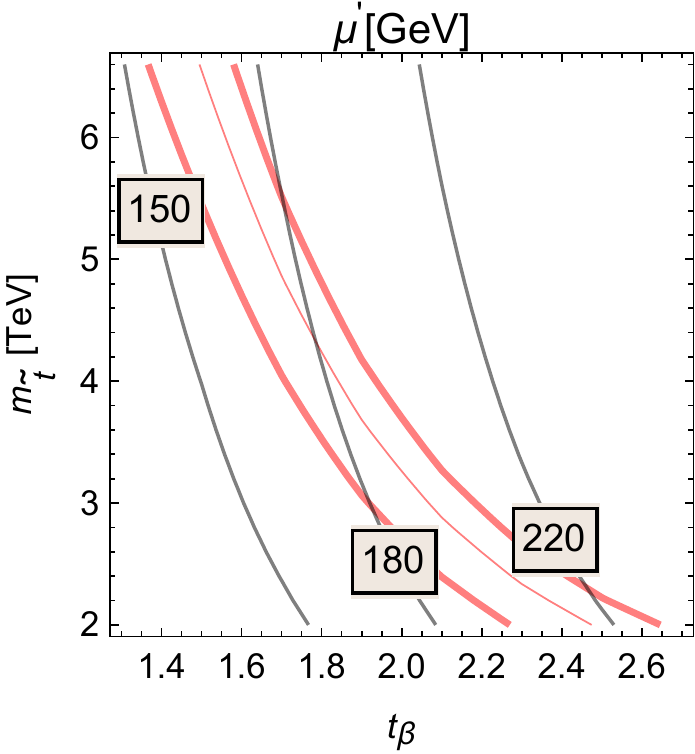}\\
\includegraphics[scale=0.75]{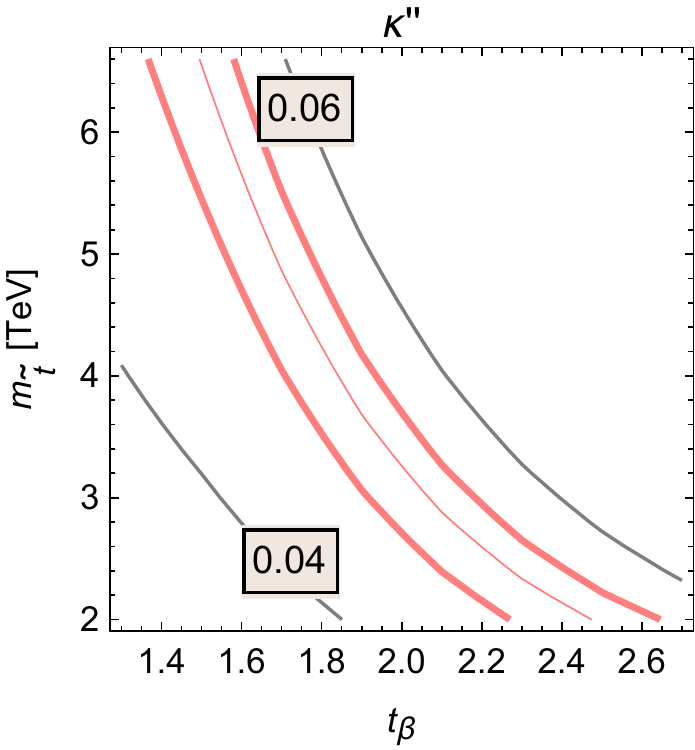}
\includegraphics[scale=0.75]{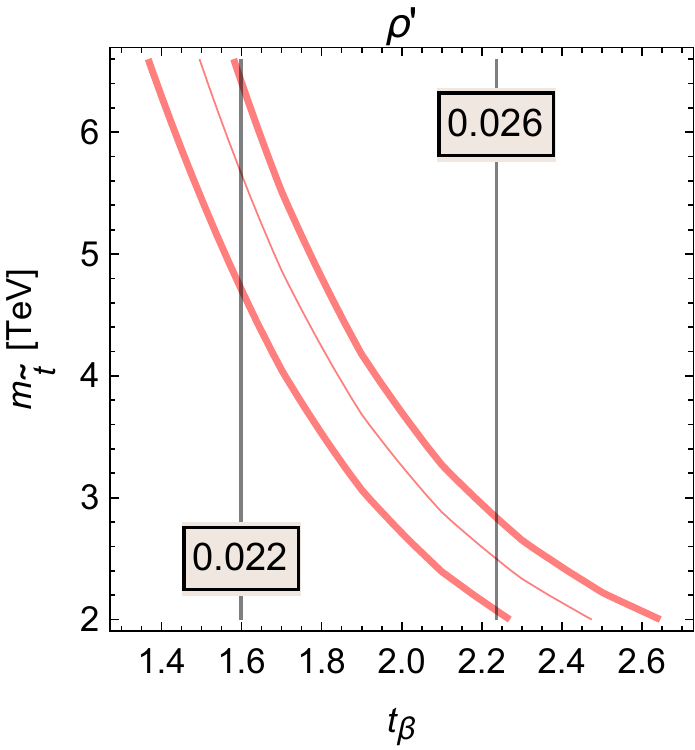}
\caption{The effective quartics and soft $\mathbb{Z}_2$ breaking term in the low-energy TMNT potential arising from a $F$-term supersymmetric UV completion with the specific values of input parameters listed in \tab{ftermpara}. The red band indicates the Higgs mass window (122 GeV, 128 GeV). The scale $F$ increases slightly with $\bar{\lambda}$ and is around 1.3 TeV everywhere.}
\label{fig:ftermpara}
\end{figure}

To illustrate the fine-tuning in the $F$-term UV completion, we fix the UV parameters as in \tab{ftermpara} and study the fine tuning in the plane $t_\beta$-$m_{\tilde{t}}$, where we translate the free parameter $m^2_u$ into the parameter $t_\beta$. The electroweak scale is fixed by the appropriate choice of $\mu^{\prime 2}$ whose value is shown in \fig{ftermpara}. Also shown in \fig{ftermpara} are couplings for the low-energy TMNT potential. The dependence of $\bar{\lambda}$ on $t_\beta$ determines that $F$ increases with $t_\beta$.  $\kappa^{\prime\prime}$ would fix the mass of the lighter mode on $H$ side in the decoupling limit and $\rho^\prime$ contributes to the SM-like Higgs quartic. The red band indicates the Higgs mass window (122 GeV, 128 GeV). 

\begin{table}
\begin{center}
\begin{tabular}{|c|c|c|c|c|c|c|c|c|}
\hline 
$m^2_d~[\mathrm{TeV^2}]$  & $\mu^2~[\mathrm{TeV^2}]$ & $b~[\mathrm{TeV^2}]$ &$m_{s0}~[\mathrm{TeV}]$& $\lambda_{s0}$  \\ 
\hline 
$0.8^2$&$0.48^2$&1.96&1.6&1.3\\ 
\hline 
$m_{sh}~[\mathrm{TeV}]$ & $m_{sH}~[\mathrm{TeV}]$ & $\mu_{sh}=\mu_{sH}~[\mathrm{TeV}]$&$\lambda_{sh}=\lambda_{sH}$&$y_t$  \\ 
\hline 
 1.0 &0.2& 0.8&1.7&0.85\\ 
\hline 
\end{tabular} 
\caption{Tree-level parameters used for our numerical study in the $F$-term supersymmetric UV completion of the TMNT model with $\mu^{\prime 2}$ solved to get $vev=\unit[174]{GeV}$. As the IR scale in our analysis is above TeV, we use $y_t=0.85$ to account for the running effects. 
}
\label{tab:ftermpara}
\end{center}
\end{table}

\begin{figure}
\centering
\includegraphics[scale=0.65]{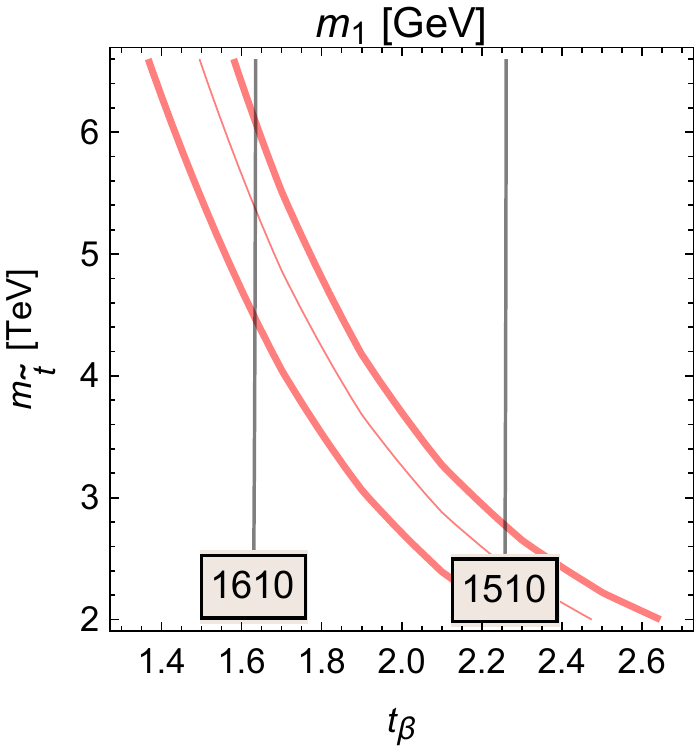}
\includegraphics[scale=0.65]{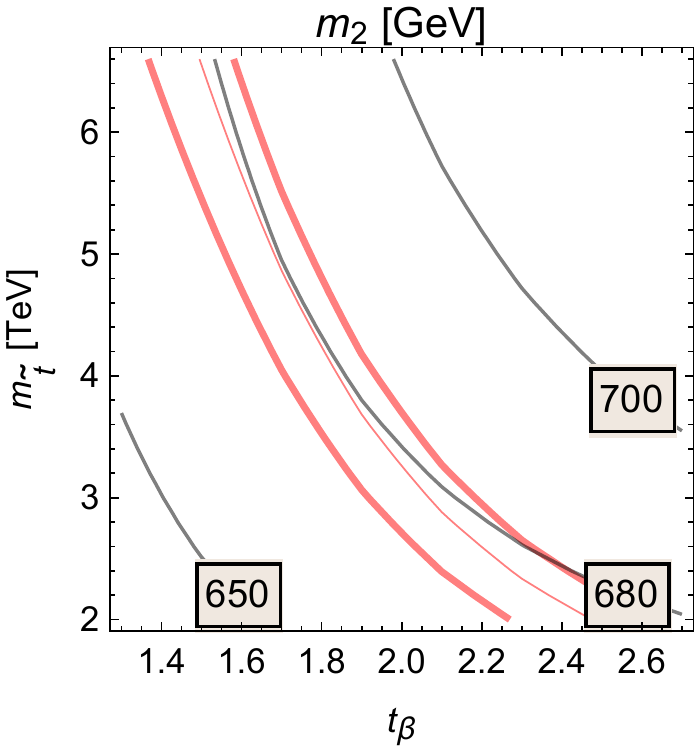}
\includegraphics[scale=0.65]{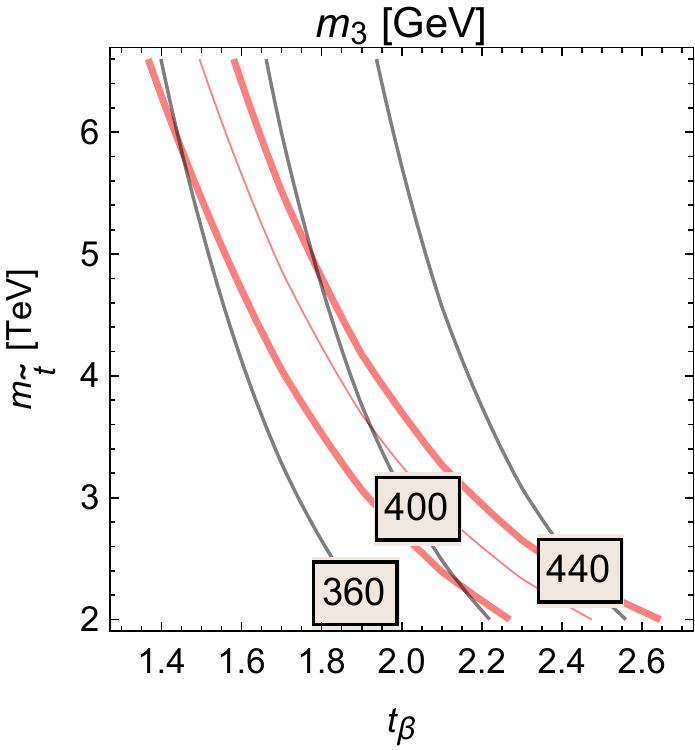}
\caption{The mass spectrum of the scalars for the numerical benchmark illustrated for the $F$-term model in this section. $m_1$ and $m_3$ are dominantly from the scalars on the $H$ side and $m_2$ is dominantly from the $h_b$ side. The red band indicates the Higgs mass window (122 GeV, 128 GeV). The scale $F$ increases slightly with $\bar{\lambda}$ and is around 1.3 TeV everywhere.}
\label{fig:ftermms}
\end{figure}

We also show the mass spectrum for the other 3 scalars in \fig{ftermms} assuming the decoupling limit. Notice that the $m_1$ mass decreases with $t_\beta$ as the symmetric mass $m$ is driven by the $b$ term and thus decreases with $t_\beta$, while $m_3$ and $m_2$ increase with $t_\beta$ because of the increasing of $\kappa^{\prime\prime}$ and the scale $f$, respectively.

\begin{figure}
\centering
\includegraphics[width=0.4\textwidth]{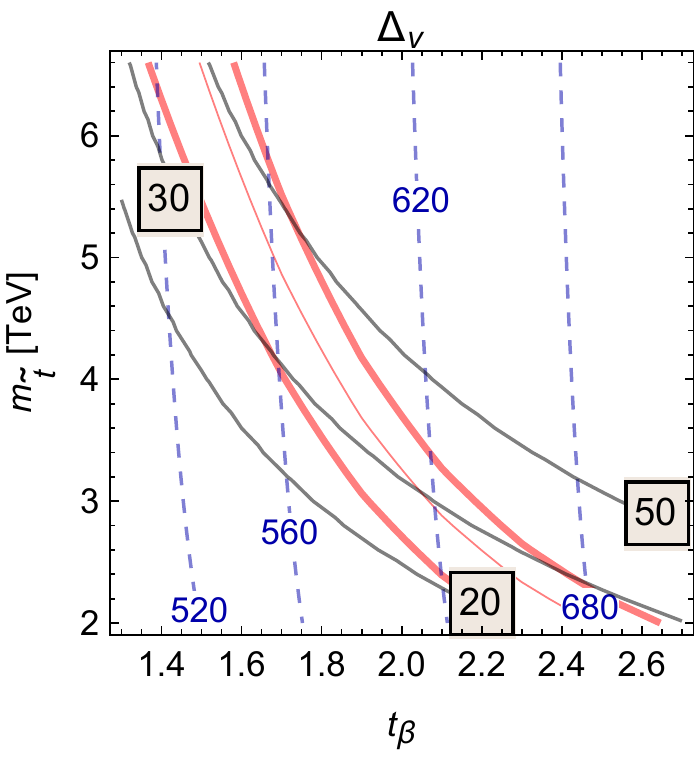}
\caption{The total tuning $\Delta_v$ of the electroweak scale $v$ in the $F$-term supersymmetric UV completion of the TMNT model with respect to all dimensional parameters added in quadrature. The red band indicates the Higgs mass window (122 GeV, 128 GeV), while the dashed line denotes the scale of $f$ in unit of GeV. The scale $F$ increases slightly with $\bar{\lambda}$ and is around 1.3 TeV everywhere.}
\label{fig:ftermtuning}
\end{figure}

The fine-tuning of the electroweak $vev$ is calculated by taking logarithmic derivatives with respect to the full set of dimensionful parameters in the UV completion, namely $P=\big\{m^2_u, m^2_d, b, \mu^2, m^2_{s0},m^2_{sh},m^2_{sH}, \mu^2_{sh}, \mu^2_{sH}, \mu^{\prime 2}, m^2_{\tilde{t}}\big\}$, and adding them in quadrature:
\begin{eqnarray}
\Delta_{v} = \sqrt{\sum_{P} \left (\frac{\partial\log v^2}{\partial \log P}\right)^2}.
\end{eqnarray}

The total fine-tuning $\Delta_v$ in the $F$-term supersymmetric completion is shown in Fig.~\ref{fig:ftermtuning}. In the non-decoupling $F$-term model, as $t_\beta$ increases, the matched quartic couplings in Eq.~\eqref{eq:matchup} decrease and thus the scales $F$ and $f$ increase. The weak dependence of the scale $f$ on $m_{\tilde{t}}$ is due to the fact that $m^2_u + \delta m^2_u$ is roughly fixed for a certain $t_\beta$ even as $m_{\tilde t}$ is varied, so that the scale $f$ is approximately constant. As such, the scale $f$ is tuned between $m^2_u$ and $\delta m^2_u$ and thus the total fine tuning $\Delta_v$ doesn't necessarily parallel the $f$ contour. Compared to the parameter space in \cite{Craig:2013fga}, we obtain the observed SM-like Higgs mass for lower values of $t_\beta$ given the same scale $f$. The $F$-term supersymmetric UV completion of the TMNT model improves the fine tuning by a factor of 3 compared to the $F$-term completion of the vanilla Twin Higgs for $m_{\tilde{t}}\sim \unit[5]{TeV}$.

%% file: topdownD.tex
In the previous section, we introduced a supersymmetric model relying on $F$-terms to generate the terms required for a UV completion of the TMNT model. We can similarly use new $U(1)$ gauge groups and their non-decoupling $D$-terms to generate the symmetry breaking terms in our model, in analogy with the $D$-term supersymmetric completion of the vanilla Twin Higgs \cite{Badziak:2017syq}. We introduce such a model, carry out the same fine-tuning analysis as above, and additionally discuss various phenomenological constraints arising from the additional gauge sectors. The overall structure is similar to Fig.~\ref{fig:setup} but with new $U(1)$ gauge groups, some singlets, and, of course, supersymmetry.

The $D$-term supersymmetric completion entails one $U(1)$ gauge group to generate the $SU(8)$ symmetric quartic and two additional $U(1)$ gauge groups to generate the $SU(4)$ symmetric quartics. We also introduce a SM singlet charged under each $U(1)$. Let's begin with the $SU(4)$ symmetric quartics. Their singlets can have the following superpotential
\begin{equation}
W \supset \lambda_{ph(H)} S_{h(H)} \left(	P_{h(H)} \bar{P}_{h(H)} - M_{ph(H)}^2	\right),
\label{eq:Wsinglets}
\end{equation}
where $S_{h(H)}$ is a singlet of $U(1)_{h(H)}$ and $P$ fields have charge $\pm q_{ph(H)}$, with the subscript in each case indicating the $SU(4)$ group of interest. These fields can have soft terms as well, which for simplicity we take to be equal for $P$ and $\bar P$:
\begin{equation}
V_{soft} \supset m^2_{ph(H)} \left(		P_{h(H)}^2 + \bar{P}_{h(H)}^2	\right).
\label{eq:Psoft}
\end{equation}
These soft terms ensure that the $D$-term quartics induced by each $U(1)$ do not decouple \cite{Batra:2003nj}. The superpotential leads to nonzero vacuum expectation values for the $P, \bar P$ fields that Higgs each of the $U(1)$s. The resulting gauge boson mass on each side is given by \cite{Batra:2003nj,Badziak:2017syq}
\begin{equation}
m_{Zh(H)} = 2 g_{h(H)} q_{ph(H)} v_{ph(H)},
\label{eq:Z'masses}
\end{equation} 
where $v_{ph(H)}$ and $g_{h(H)}$ denote the \textit{vev}s and the gauge couplings of each side, respectively. As in the $F$-term case, we can match on to the low-energy TMNT model by rotating to the Higgs basis, thereby isolating contributions to the light scalars that remain in the decoupling limit of the supersymmetric 2HDMs. Assigning a charge of $q=1/2$ to $h$ and $H$ fields under their respective $U(1)$s, the non-decoupling $D$-term quartics on each side are
\begin{eqnarray}
\label{eq:nonDhH}
V_{Dh} &\supset & \frac{g_h^2}{8} h^4 \cos^2 (2\beta_h) \left(\frac{2m_{ph}^2}{2m_{ph}^2+m_{Zh}^2}\right),  \\ 
V_{DH} &\supset & \frac{g_H^2}{8} H^4 \cos^2 (2\beta_H) \left(\frac{2m_{pH}^2}{2m_{pH}^2+m_{ZH}^2}\right).\nonumber
\end{eqnarray}
For the rest of our study, we will concentrate on equal $\beta$ ($=\beta_h=\beta_H$) angle on each side. 

We repeat the exercise for the $SU(8)$ symmetric quartic, introducing an analogous set of multiplets; we denote these with the subscript ``0''. The quartic resulting from this $U(1)$ is simply
\begin{eqnarray}
\label{eq:nonD}
V_{D} &\supset & \frac{g_0^2}{8} \left( h^2  + H^2  \right)^2 \cos^2 (2\beta)\left(\frac{2m_{p0}^2}{2m_{p0}^2+m_{Z0}^2}\right).
\end{eqnarray}

The $D$-term in Eq.~\eqref{eq:nonD} can be mapped to the coupling $\bar{\lambda}$ in the general potential in Eq.~\eqref{eq:fullpot}, while those in Eq.~\eqref{eq:nonDhH} generate $\bar{\kappa}$ and $\bar{\rho}$. In particular, $\bar \rho$ can be generated by taking either the soft scales $m_{ph}^2, m_{pH}^2$ to be unequal, or the supersymmetric scales $m_{Zh}^2, m_{ZH}^2$ to be unequal, thereby generating an effective hard-breaking quartic via soft terms.

To simplify our analysis, we assume the new singlet \textit{vev}s and the gauge couplings at $SU(4)$ level are equal at different sides. Hence, the $\mathbb{Z}_2$-symmetry breaking is sourced only through unequal soft terms. Assuming $g_H=g_h$, the generated quartics are then given by
\begin{eqnarray}
\bar{\lambda} & = & \frac{g_0^2}{8} \frac{2m_{p0}^2}{2m_{p0}^2+m_{Z0}^2} \cos^2 (2\beta), \nonumber \\
\label{eq:lambdas127}
\bar{\kappa} & = & \frac{g_h^2}{8} \frac{2m_{pH}^2}{2m_{pH}^2+m_{Zh}^2} \cos^2 (2\beta), \\
\bar{\rho} & = & \frac{g_h^2}{8} \left(\frac{m_{ZH}^2}{2m_{pH}^2+m_{Zh}^2} - \frac{m_{Zh}^2}{2m_{ph}^2+m_{Zh}^2} \right) \cos^2 (2\beta). \nonumber 
\end{eqnarray}

The symmetric mass of the general potential at this level can arise from the usual mass terms in a SUSY model, namely $m_u,m_d,\mu^2,$ and $b$, as in the $F$-term model. We assume these terms are the same on each side so they only contribute to the totally symmetric mass term at tree-level\footnote{Through loop corrections, the $\mathbb{Z}_2$--breaking terms do generate some misalignment in these parameters between different sites.}. In what follows, we will work with the tree-level expressions above for the respective quartics, as the loop corrections will be proportional to these quantities as well and are suppressed compared to the leading pieces. 

The SM gauge group (and its twin) make an additional $D$-term contribution to the quartics at tree-level, while top loops give an appreciable one-loop contribution as well. These corrections have the form
\begin{eqnarray}
\kappa'' & = &\frac{g_{ew,twin}^2}{8} \cos^2(2\beta) + \frac{3 y_t^4}{16\pi^2} \log \left(	\frac{m_{\tilde{t}}^2}{m_{T_a}^2}	\right),\nonumber \\
\label{eq:lambdas356}
\kappa' & = &\frac{g_{ew}^2}{8} \cos^2(2\beta) + \frac{3 y_t^4}{16\pi^2} \log \left(	\frac{m_{\tilde{t}}^2}{m_{t_b}^2}	\right) - \kappa'', \\
\rho' & = & \frac{3 y_t^4}{16\pi^2} \log \left(	\frac{f^2}{v^2}	\right) \nonumber ,
\end{eqnarray} 
where we will use $g_{ew,twin}=g_{ew}$.

We include the correction to the soft terms in the general potential from the loops of the new gauge bosons and the stops loops. In contrast to the $F$-term model, where corrections from the singlet sector soft mass terms have a logarithmic dependence on the messenger scale $\Lambda_{mess}$, the corrections from the new gauge sector soft mass terms do not depend on $\Lambda_{mess}$ at one loop. Hence, the loop corrections from the scalar and fermionic partners of the $h$ supermultiplets will not be sub-dominant anymore. We include the contribution of these loops to $m'^2_h$. Due to supersymmetry, the quadratic pieces of these fermionic and bosonic loops will cancel; the log piece, however, gives rise to an analog of Eq.~\eqref{eq:epm}, where instead of a quadratic cutoff we will have the soft mass of the scalar component of the $h$ supermultiplet. Collectively, these effects contribute to the TMNT mass parameters as
\begin{eqnarray}
\label{eq:softsg}
\delta m^2& = & \frac{g_0^2}{64\pi^2} m_{Z0}^2 \log \left(	\frac{2m_{p0}^2+m_{Z0}^2}{m_{Z0}^2} \right) + \frac{g_h^2}{64\pi^2} m_{ZH}^2 \log \left(	\frac{2m_{pH}^2+m_{ZH}^2}{m_{ZH}^2}	\right) -\frac{3 y^2_t}{16\pi^2} m_{\tilde{t}}^2 \log\frac{\Lambda_{mess}^2}{m_{\tilde{t}}^2}, \nonumber \\
\delta m_h'^2 & = & \frac{5\bar{\rho}}{16\pi^2} \tilde{\Lambda}_\rho^2+ 	\frac{g_h^2}{64\pi^2} m_{Zh}^2 \log \left(		\frac{2m_{ph}^2+m_{Zh}^2}{2m_{pH}^2+m_{ZH}^2}	 \right)		 , 
\end{eqnarray}
where in the decoupling limit
\begin{equation}
\tilde{\Lambda}_\rho^2 \sim - m_u^2 \log \left(		\frac{m_{ph}^2}{M_{h}^2}		\right) ,
\label{eq:Lambdarhot}
\end{equation}
$M_h$ denotes the heavy Higgs mass of the $h$ field that is decoupled, $m_{ph}$ is the mass of the $P$ singlet mass, the factor of 5 in the second expression in Eq.~\eqref{eq:softsg} is due to the scalar multiplicity, and similar to the previous section we assume the messenger scale is at $\Lambda_{mess}=10m_{\tilde{t}}$. Note should be taken that even though $\tilde{\Lambda}_\rho$ emulates a cutoff as in Eq.~\eqref{eq:epm}, it is in fact a log correction. To stabilize the \textit{vev}s in our setup we have $m_u^2<0$, thus $\tilde{\Lambda}_\rho^2 >0$. 

We have not introduced any source for generating the soft $\mathbb{Z}_2$-breaking $\mu'^2$ term. We assume it is generated by another mechanism not included here and determine its value numerically such that $v$ has the correct numerical value.

\subsubsection{Constraints \& Fine Tuning}

As with the $F$-term model, we now turn to the tuning of the electroweak scale in the $D$-term scenario. The viable parameter space is subject to somewhat stronger experimental constraints due to the effects of the additional $U(1)$ gauge groups, which we must accommodate before determining the numerical fine-tuning. In particular, we must not only consider the constraints imposed by avoiding low-scale Landau poles associated with the new gauge groups, but also bounds coming from electroweak precision observables (in particular oblique parameters $S$ and $T$ \cite{Peskin:1990zt,Peskin:1991sw,Cacciapaglia:2006pk}) due to Higgs mixing and the new $Z^\prime$ gauge bosons.

Let us begin with the Landau poles. We are introducing three new $U(1)$ gauge groups, each of whose gauge couplings can develop a Landau pole in the UV. As we will focus on large gauge couplings (to generate sizable quartics), this threat arises at very low scales, i.e. below 1000~TeV. Avoiding these low-scale Landau poles requires further UV-completion of these gauge groups, e.g. embedding them inside a larger group. We postpone further study of such UV completions for now and merely demand the Landau poles to be above a fiducial value, say 100~TeV. 

The one-loop Landau pole for a particular $U(1)$ gauge group is given by 
\begin{equation}
\Lambda_{L} = \Lambda_{IR} \exp \left(		\frac{2\pi}{A \alpha_{IR}}		\right),
\label{eq:landau}
\end{equation} 
where $A$ stands for the quadrature sum of field charges under that group. We assume only the Higgs supermultiplets (and not other new or SM matter content) are charged under each gauge group, all with charge $1/2$\footnote{By not charging SM field content under these new gauges, the Standard Model yukawas are necessarily generated by some non-renormalizable operators suppressed by a UV scale. This could explain the small value and the hierarchy of the SM yukawa couplings; see, for example, \cite{Craig:2011yk} for a similar idea in the context of deconstruction models \cite{ArkaniHamed:2001nc}.}. 
\begin{figure}
\begin{center}
\includegraphics[scale=0.65]{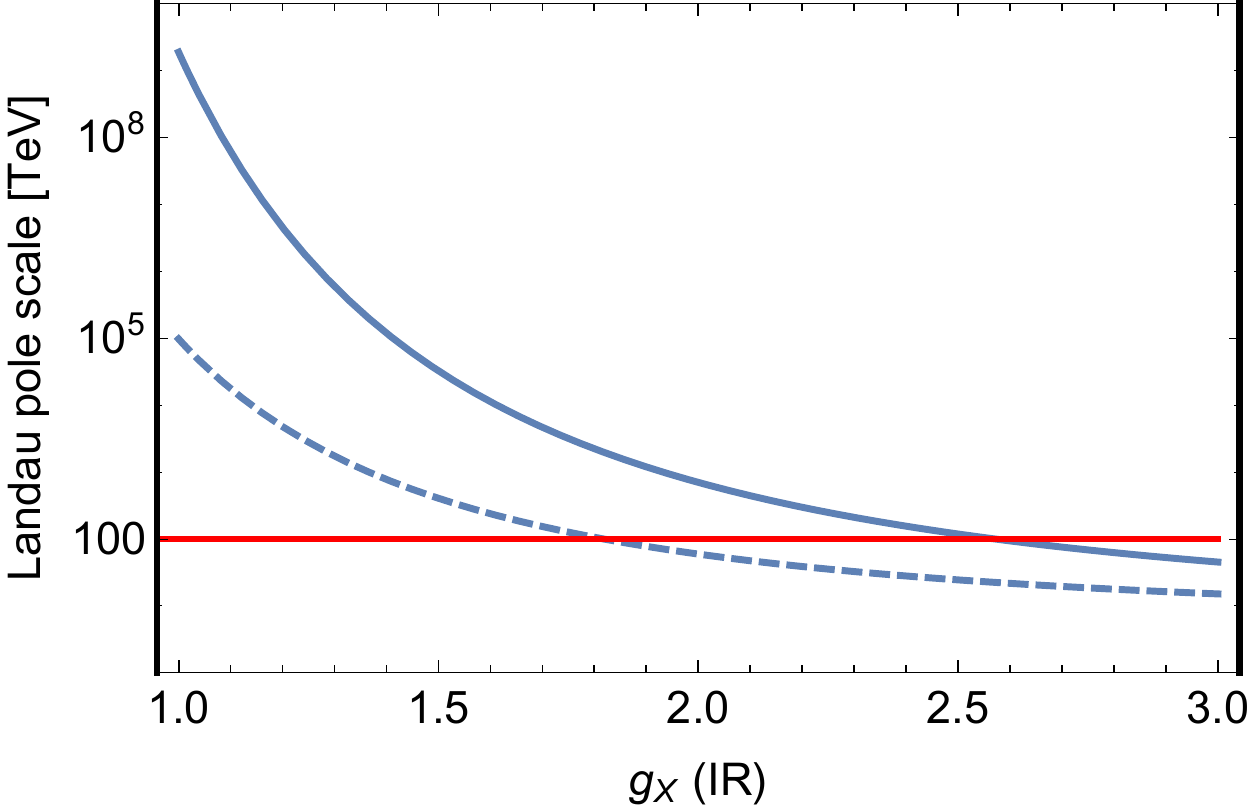}
\caption{Location of the one-loop Landau pole for the $U(1)$ gauge groups associated with the $SU(4)$ symmetric quartics (solid line) and the gauge group associated with the $SU(8)$ symmetric quartic (dashed line). Here we assume only the supermultiplets containing Higgs-like fields are charged under the new groups, and the IR scale is taken to be $\sim5$TeV. Further UV completion is required at or below the scale of these poles. Demanding the Landau poles to be above a certain value, e.g. 100~TeV, limits the size of the $D$-term quartics in our model.}
\label{fig:landau}
\end{center}
\end{figure}
The Landau poles as a function of the IR gauge couplings (taken at 5~TeV) are shown in Fig.~\ref{fig:landau}. This illustrates the upper bound on the gauge couplings implied by avoiding a Landau pole beneath a certain scale. In our numerical studies we will fix $g_0=1.6$ to have a Landau pole above $100$~TeV, and take $g_h=1.4$.

The most pressing experimental constraints arise from precision electroweak limits; here we focus on $S$ and $T$. The contributions to $S$ and $T$ from the mixing of the SM-like Higgs with other sectors can be made sufficiently small for large enough $f$. More pressing is the contribution from the new $Z^\prime$ gauge bosons, which contribute to the $T$ parameter at tree level with the charge assignments articulated above. Using the current bounds on this parameter \cite{PhysRevD.98.030001} and the formalism of \cite{Cacciapaglia:2006pk}, we find that staying within $1 \sigma$ of the current central value implies
\begin{equation}
\frac{m_{Z'}}{g_{Z'}} \sim v_p \gtrsim 3\mathrm{TeV}.
\label{eq:vprange}
\end{equation}
In what follows, we assume the \textit{vev}s $v_p$ are a factor of $\sqrt{2}$ greater than the bound above, to accommodate the fact that there are multiple vectors contributing to $T$.

We are now in a position to consider the numerical fine-tuning of the weak scale, following the prescription detailed in Sec.~\ref{sec:tuning}. We add the contributions from all of the dimensionful underlying parameters in quadrature. In this model, these parameters are
\begin{equation}
\lbrace	v_p^2(=v_{ph}^2=v_{pH}^2), v_{p0}^2, m_{ph}^2, m_{pH}^2, m_{p0}^2, m_{u}^2,m_{d}^2, \mu^2, b, \mu'^2	,m_{\tilde{t}}^2 \rbrace.
\label{eq:FTparams}
\end{equation}

\begin{figure}
\begin{center}
\includegraphics[scale=0.75]{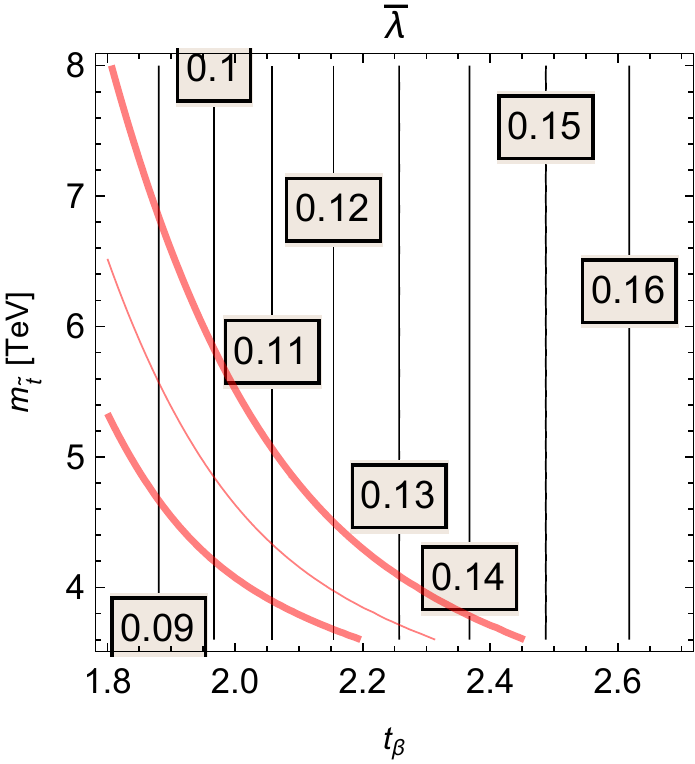}
\includegraphics[scale=0.75]{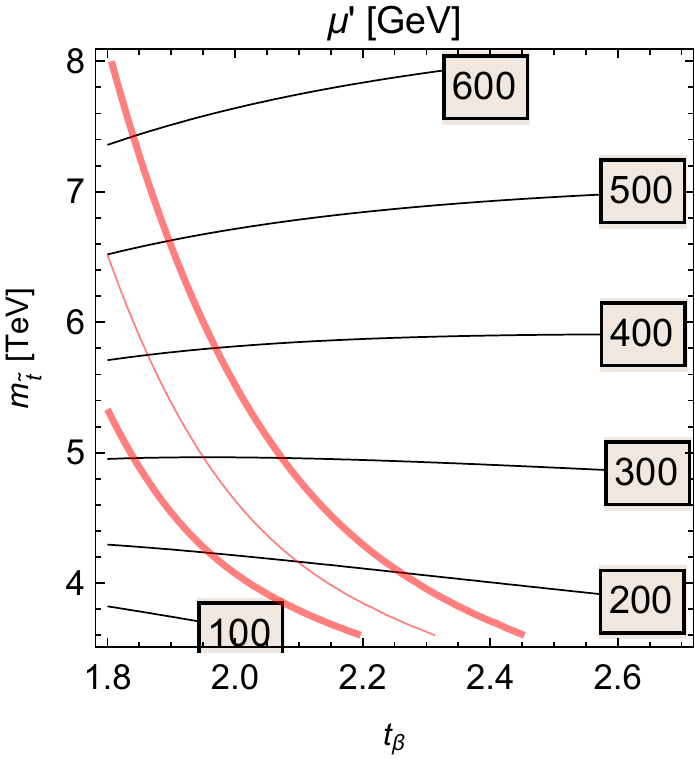}\\
\includegraphics[scale=0.75]{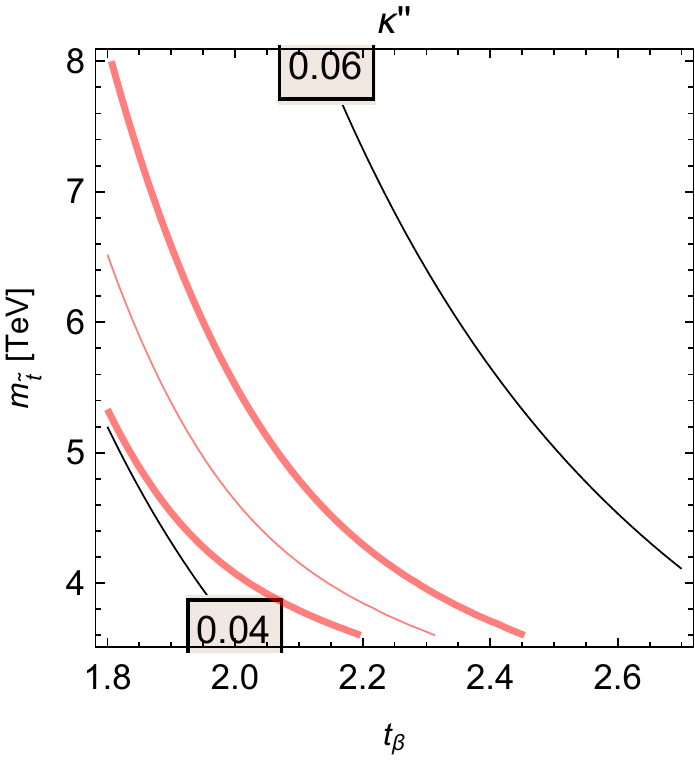}
\includegraphics[scale=0.75]{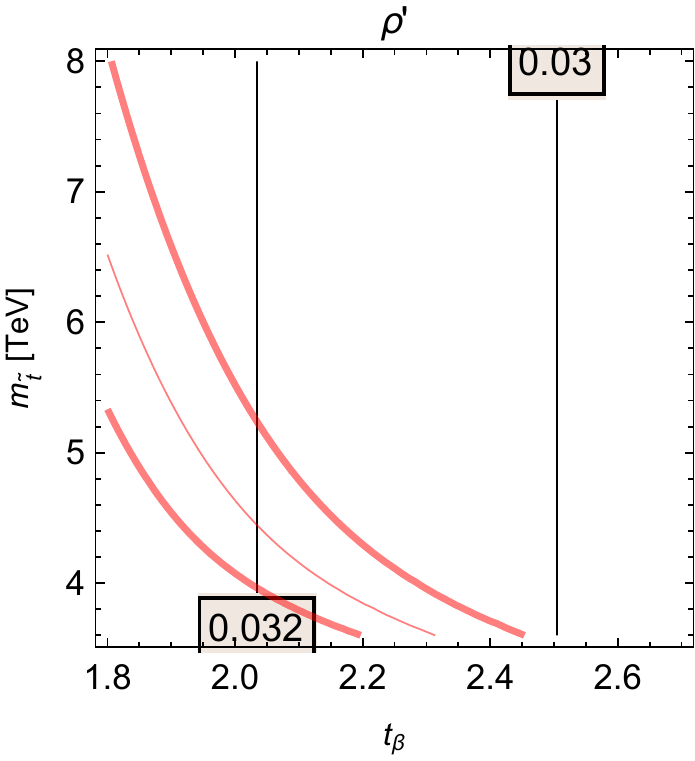}
\caption{The effective quartics and soft $\mathbb{Z}_2$ breaking term in the low-energy TMNT potential arising from a $D$-term supersymmetric UV completion with the specific values of input parameters listed in Table~\ref{tab:numbers}. The red lines in this figure indicate the range of parameters that generates a higgs mass between $\left(		122 , 128		\right)$GeV. }
\label{fig:couplingsD}
\end{center}
\end{figure}

As there are many underlying parameters, to demonstrate the qualitative features of fine-tuning we fix many of them to particular numerical values and illustrate the tuning as a function of two parameters. Other underlying parameters are fixed to numerical values indicated in Table~\ref{tab:numbers}. The couplings in the low-energy TMNT potential, derived from these input parameters and  as a function of the cutoff scale $m_{\tilde{t}}$ and $\tan \beta$ ($t_\beta$), are illustrated in Fig.~\ref{fig:couplingsD}. Notice the different $t_\beta$ dependence of $\bar{\lambda}$ coupling compared to the $F$-term model; as a result, in this model the scales $f$ and $F$ have a different dependence on $t_\beta$, which consequently gives rise to a different behavior for the electroweak scale tuning in this model. These figures also indicate the dependence of $\kappa''$ and $\rho'$ on $\beta$ and a slight dependence on $m_{\tilde{t}}$ due to the log factor in Eq.~\eqref{eq:lambdas356}. The red lines in this figure indicate the range of parameters that generates a higgs mass between $\left(		122 , 128		\right)$ GeV. 

\begin{figure}
\begin{center}
\includegraphics[scale=0.65]{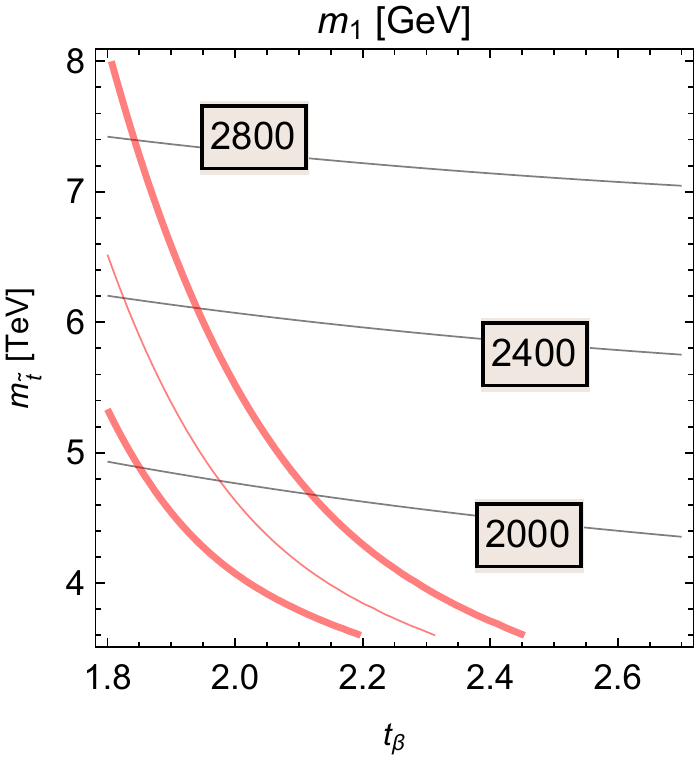}
\includegraphics[scale=0.65]{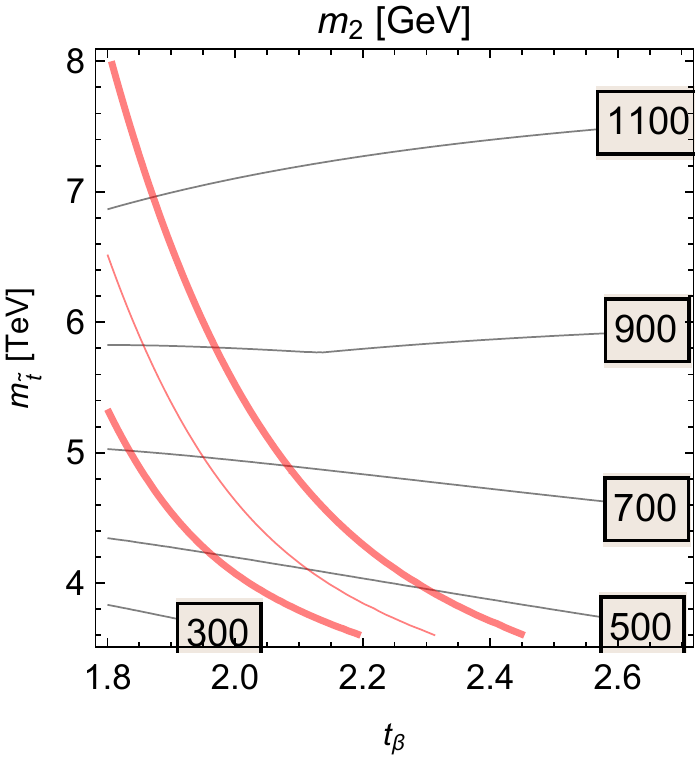}
\includegraphics[scale=0.65]{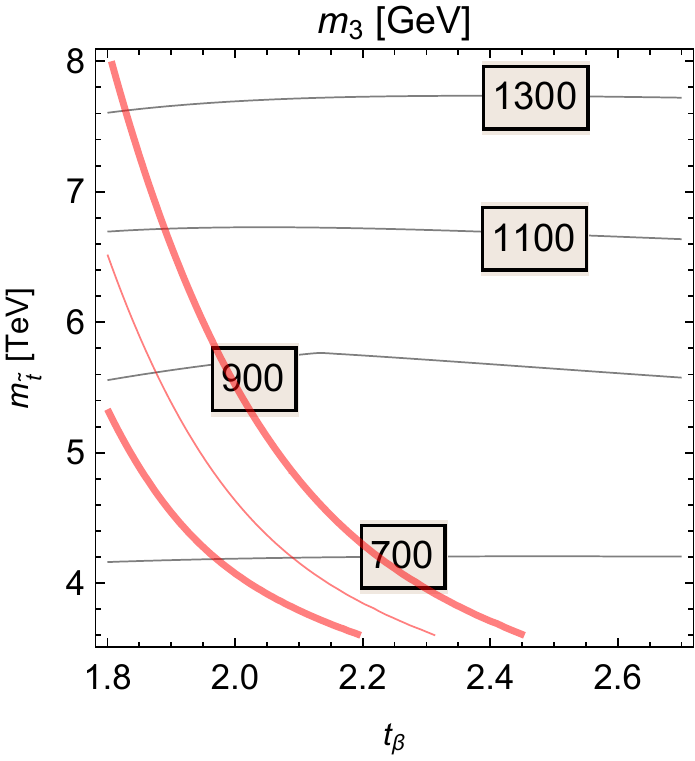}
\caption{The mass spectrum of the scalars for the numerical benchmark illustrated in this section. $m_1$ is dominantly from the scalars on the $H$ side of the tower. Notice the level-crossing between $m_2$ and $m_3$ in the 900~GeV contour; one of these eigenvalues is mostly comprised of $h_b$, while the other one is mainly made of the $H$ fields. }
\label{fig:massesD}
\end{center}
\end{figure}

Assuming the decoupling limit, there are 3 other scalar masses in the spectrum. These eigenvalues are shown in Fig.~\ref{fig:massesD}. We observe a level-crossing between two of the eigenvalues in these plots. For almost all the range of parameters we are studying the lightest eigenvalue (the SM-like Higgs) has small enough mixing with other sectors to evade the electroweak bounds due to its mixing with the twin sector \cite{Craig:2013fga,Katz:2016wtw}.

\begin{table}
\begin{center}
\begin{tabular}{|c|c|c|c|c|}
\hline 
$g_0$ & $g_h=g_H$ & $v_p=v_{p0}~\mathrm{TeV}$ & $m_{ph}~[\mathrm{TeV}]$ & $m_{pH}~[\mathrm{TeV}]$ \\ 
\hline 
1.6 & 1.4 & 4.5 & 14.3 & 2.25 \\ 
\hline 
$m_{p0}~[\mathrm{TeV}]$ & $m^2_{u}~[\mathrm{TeV}^2]$ & $m^2_{d}~[\mathrm{TeV}^2]$ & $\mu^2~[\mathrm{TeV}^2]$ & $y_t$ \\ 
\hline 
14.3 & $-1.4^2$ & $-0.1^2$ & $0.48^2$ & 0.85 \\ 
\hline 
\end{tabular}
\caption{Numerical values of the input parameters used in our study. As the IR scale in our analysis is above a TeV, we use a smaller value of $y_t$ than its weak-scale value to account for running effects. The only source of $\mathbb{Z}_2$ breaking between the two sides of our tower are the soft masses of the $P$ fields.}
\label{tab:numbers} 
\end{center}
\end{table}

\begin{figure}
\begin{center}
\includegraphics[scale=0.85]{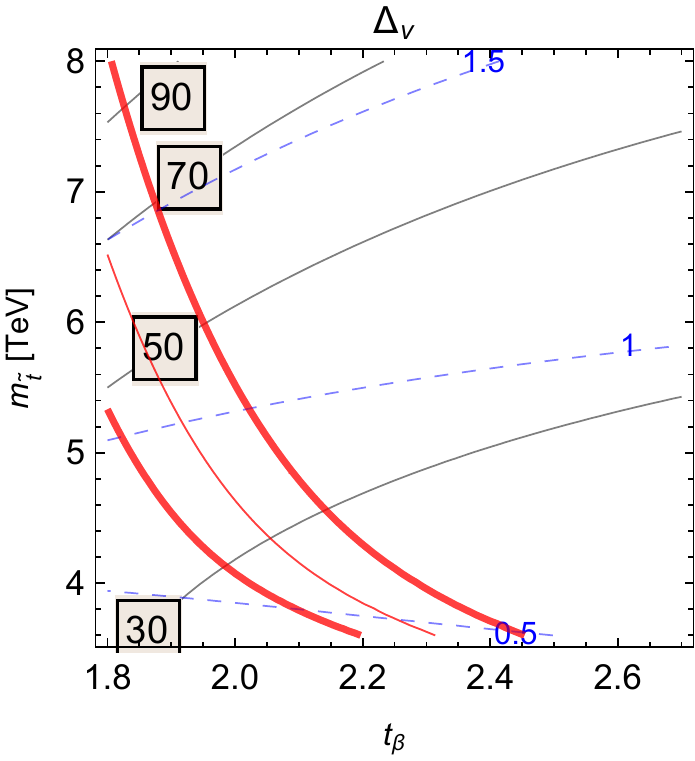}
\caption{The total fine-tuning of the electroweak scale $v$ with respect to all the underlying parameters in the $D$-term model. We also show the values of the \textit{vev} $f$ in unit of TeV with blue dashed lines. The numerical benchmark values used for this plot are included in Table~\ref{tab:numbers}. The contours of constant tuning have a different behavior compared to the $F$-term model in Fig.~\ref{fig:ftermtuning}. This difference can be mainly attributed to the dependence of the symmetric coupling $\bar{\lambda}$ on the $\beta$ angle in the two models. }
\label{fig:tuningD}
\end{center}
\end{figure}

The total tuning of the electroweak scale with respect to all the parameters in our model is shown in Fig.~\ref{fig:tuningD}. The values of the \textit{vev} $f$ are also shown in blue dashed lines. The scale $F$ behaves similar to the $f$ \textit{vev}; for the parameters in Table~\ref{tab:numbers} and the range of $m_{\tilde{t}}$ and $\tan \beta$ we are studying, $F$ varies between $2-4$TeV. Figure \ref{fig:tuningD} indicates similar fine-tuning improvement to the $F$-term model, which is improved by a factor of 3-4 compared to a generic supersymmetric Twin Higgs model. As argued earlier, the TMNT gain in fine-tuning and its ability to raise the scale of new colored physics (denoted by the stop mass $m_{\tilde{t}}$) comes at the cost of additional Higgs scalars and a relatively low scale of UV completion for the Higgs quartic.

In contrast to the $F$-term model in Sec.~\ref{subsec:fterm}, the symmetric quartic in this model increases with $\tan \beta$. As a result, as we go to larger $\tan \beta$ the scale $f$ decreases as indicated in Fig.~\ref{fig:tuningD}. This, in turn, suggests an improvement in the tuning as we go to larger $\tan \beta$. Using the factorization arguments for the tuning in TMNT in Sec.~\ref{sec:bottomup}, the dominant tuning is between the scales $f$ and $v$, thus the contours of the tuning, to a good extent, follow the constant $f$ contours. As the $m_{\tilde{t}}$ scale increases, the fine-tuning should naturally worsen. These effects give rise to the behavior of the fine-tuning as a function of $\tan \beta - m_{\tilde{t}}$ in Fig.~\ref{fig:tuningD}. Notice the different behavior of contours of constant tuning compared to the $F$-term supersymmetric models in the previous section; as explained, this difference should be attributed to the dependence of the symmetric quartic $\bar{\lambda}$ and, consequently, the \textit{vev}s $f$ and $F$ on $\tan \beta$.

Numerical investigation shows that a dominant source of tuning in our model is the $\mu'^2$ term in the downstairs theory. This is indeed in accordance with our factorization analysis in Sec.~\ref{sec:bottomup} that suggests the softly-broken downstairs $\mathbb{Z}_2$ symmetry is the prevailing source of tuning in the TMNT. 

While the overall tuning is worsened in these UV completions, this is a common feature of all supersymmetric UV completions of the Twin Higgs (e.g.~\cite{Craig:2013fga,Katz:2016wtw}). 
Crucially, however, the UV completions validate the parametric improvement in tuning suggested by the analysis of the effective theory, in the sense that they improve tuning compared to supersymmetric UV completions of the vanilla Twin Higgs. Moreover, these supersymmetric UV completions validate some of the parametric choices made in constructing the bottom-up TMNT model, demonstrating that they can be preserved by more complete theories.

%% file: conclusion.tex
In this work, we have extended the framework of neutral naturalness by demonstrating that the hierarchical breaking of discrete symmetries can stabilize the Higgs mass and further increase the natural scale of new colored states associated with compositeness or supersymmetry as an ultimate UV completion. The increased scale of colored states is obtained at the price of introducing additional Higgs-like scalars that couple weakly to the Standard Model, making additional Higgs-like scalars the most immediate signature of naturalness in this framework. Additionally, the scale associated with new electroweak-charged states responsible for the UV completion of quartic couplings is roughly the same as in the Standard Model or conventional Twin Higgs models. We have not only illustrated the mechanism in terms of an effective extension of the Standard Model up to some further cutoff, but also provided two explicit supersymmetric UV completions that validate the expectations of the effective theory. Along the way, we have uncovered a novel mechanism for UV completing hard $\mathbb{Z}_2$-breaking quartics in Twin Higgs models via soft supersymmetry breaking masses, which may be fruitfully applied to UV completions of more conventional Twin Higgs models as well.

There are many possible opportunities and advantages in such `turtle' constructions, of which the framework presented here is only one example. One particularly promising direction for further exploration is the construction of turtle models based on discrete symmetries that push off all divergences to as many loops as possible, in the spirit of the original turtle models. A useful starting-point for such models might be \cite{Chacko:2005vw}, in which a collectively-generated tree-level $\mathbb{Z}_2$-symmetric quartic term allows for the separation of $v$ from $f$ via soft $\mathbb{Z}_2$ breaking without corresponding $\mathcal{O}(f^2/v^2)$ tuning. A turtle model based on this framework would allow for the separation of intermediate scales of discrete symmetry breaking without additional tuning {\it or} the reintroduction of quadratic sensitivity to the cutoff. 

Another promising direction is the construction of other patterns of discrete symmetry breaking. In this work we have constructed what is in some sense the most naive turtle extension of the Twin Higgs by squaring the underlying $SU(2) \times SU(2) \times \mathbb{Z}_2$ structure. But other patterns of discrete symmetry may yield greater simplicity and further improvement in tuning. More broadly, given the pressing need for innovation in our approach to the electroweak hierarchy problem, the further exploration of turtle constructions in general is likely to bear significant fruit.